\documentclass[preprint,review,12pt]{elsarticle}

\usepackage[T1]{fontenc}
\usepackage{natbib}
\usepackage{float}
\usepackage{bm}
\usepackage{siunitx}
\usepackage{geometry}
\usepackage{tikz}
\usepackage{fontawesome}
\usepackage{tabularx}
\usetikzlibrary{patterns}
\usepackage{amsmath, graphicx,amssymb,epsfig,psfrag,subfigure,bm,bbm,booktabs}
\usepackage{mathrsfs}
\usepackage{cases}
\usepackage{multirow} % for cmd 'multirow', 'multicolumn'
\usepackage{color}
\usepackage{hyperref}
\hypersetup{hypertex=true,
	colorlinks=true,
	linkcolor=blue,
	anchorcolor=blue,
	citecolor=blue}

\newcommand{\bpm}{\begin{pmatrix}}
\newcommand{\epm}{\end{pmatrix}}
\newcommand{\beqa}{\begin{eqnarray}}
\newcommand{\eeqa}{\end{eqnarray}}
\newcommand{\beqas}{\begin{eqnarray*}}
\newcommand{\eeqas}{\end{eqnarray*}}

\newcommand{\rT}{\mathrm{T}}

\newcommand{\rb}{\mathrm{b}}

\newcommand{\mR}{\mathbb{R}}

\newcommand{\CL}{\mathcal{L}}

\newcommand{\CF}{\mathcal{F}}

\newcommand{\vx}{\mathbf{x}}
\newcommand{\vy}{\mathbf{y}}
\newcommand{\vz}{\mathbf{z}}

\newcommand{\vn}{\mathbf{n}}
\newcommand{\vv}{\mathbf{v}}
\newcommand{\vg}{\mathbf{g}}
\newcommand{\vt}{\mathbf{t}}
\newcommand{\vu}{\mathbf{u}}
\newcommand{\vq}{\bm{q}}
\newcommand{\vd}{\bm{d}}
\newcommand{\va}{\bm{a}}
\newcommand{\ve}{\bm{e}}

\newcommand{\vc}{\bm{c}}

\newcommand{\vh}{\bm{h}}

\newcommand{\vC}{\bm{C}}

\newcommand{\vS}{\bm{S}}
\newcommand{\vX}{\bm{X}}
\newcommand{\vF}{\bm{F}}

\newcommand{\vN}{\bm{N}}
\newcommand{\vP}{\bm{P}}

\newcommand{\vQ}{\bm{Q}}
\newcommand{\vR}{\bm{R}}

\newcommand{\vbeta}{\boldsymbol{\beta}}
\newcommand{\vgamma}{\boldsymbol{\gamma}}

\newcommand{\mO}{\mathcal{O}}

\newcommand{\rd}{\mathrm{d}}
\newcolumntype{Z}{>{\centering\let\newline\\\arraybackslash\hspace{0pt}}X}

\makeatletter
\newcommand*\bigcdot{\mathpalette\bigcdot@{.5}}
\newcommand*\bigcdot@[2]{\mathbin{\vcenter{\hbox{\scalebox{#2}{$\m@th#1\bullet$}}}}}
\makeatother

\journal{}

\begin{document}

\begin{frontmatter}



\title{A neural accelerated roughness-resolved Discrete Element Method (Rough-DEM) for modeling granular systems}

\author[1]{Xiwei Pan}
\ead{xp9583@princeton.edu}

\author[1,2]{Liuchi Li\corref{cor1}}
\ead{ll2756@princeton.edu}

\address[1]{Department of Civil and Environmental Engineering, Princeton University, 08544, USA}
\address[2]{Princeton Materials Institute, Princeton University, 08544, USA}

\cortext[cor1]{Corresponding author}

\begin{abstract}
The Discrete Element Method (DEM) is an indispensable tool for modeling and understanding the basic mechanical behaviors of (dry) granular media, a class of disordered material systems of significant interest to multiple scientific domains from (geo-)physics to (chemical) engineering. At the core of all present DEM frameworks is a phenomenological friction law—commonly chosen to be of the Amontons-Coulomb type—that models the frictional interaction among constituent grains. Here, inspired by research in the tribology community, we present an alternative, fundamentally different framework, termed roughness-resolved DEM (Rough-DEM), in which contact friction is handled in a more physics-based manner. In this framework, instead of using a friction law, we explicitly resolve the surface topography of each grain and let friction emerge from collective frictionless interactions of surface asperities. In essence, we trade phenomenology for physics at the cost of an increased computational demand. To mitigate the cost, we introduce an FE$^2$-like algorithm and a feature-wise linear modulation (FiLM)-based neural network surrogate, both of which are parallelized on GPU to accelerate contact computation. Crucially, using this framework in 2D, we (1) demonstrate the possibility of recovering a two-body-based Amontons-Coulomb-type friction behavior using a \textit{simple} representation of surface topography, and (2) identify many-body-based frictional behaviors—more specifically, the angle of repose values—that \textit{cannot always} be captured by current DEM frameworks under the same level of model complexity. We conclude by discussing future directions for Rough-DEM, as well as exciting new opportunities it could unlock for a wide range of emerging applications, including (designing) suspension rheology and metamaterial mechanics.
\end{abstract}


\end{frontmatter}


\section{Introduction}
\label{sec:introduction}

The Discrete Element Method (DEM)~\cite{cundall1979discrete} has quickly evolved—over the years since its inception in 1979—to be the standard numerical tool for modeling and understanding the basic mechanical behaviors of (dry) granular media, particularly in scenarios where their macroscopic dynamics are dominated by collective grain-scale rearrangements rather than grain-scale mechanical deformations (i.e., ``rigid'' grains). Its applications span multiple scientific domains: from fault gouge~\cite{ferdowsi2014three,papachristos2023discrete} and rupture~\cite{garcia2019discrete} in (geo-)physics to mixing in (chemical) engineering~\cite{fry2019diffusion}, and even to locomotion in robotics~\cite{zhang2013ground,nazemi2026reciprocal}, just to name a few. For these applications, when needed, DEM realizations are treated as the ``ground truths'' in lieu of experiments to validate reduced-order theories in describing a granular system's macroscopic behavior; see ~\cite{kamrin2012nonlocal,xiao2016modelling,slonaker2017general,kim2020power,sharma2026gravity} for a few examples.

The widespread popularity of DEM—similar to that of Molecular Dynamics (MD)~\cite{alder1957phase}—lies in its distinct ability to access high-resolution data: time-resolved kinematics (e.g., acceleration, velocity, and position) of every single grain and contact forces in a system, both of which are otherwise difficult (if not completely impossible) to get from experiments, albeit recent advances in optical diagnostic tools~\cite{hurley2016quantifying,hurley2023assessing,li2024dynamic}; and this is especially true for large systems (i.e., with over a few thousands grains)—a scenario naturally more amenable to DEM given rapid advances in computing capacities, both software-wise (e.g., parallel computing and machine learning) and hardware-wise (e.g., GPU and RAM) speaking. Through the lens of mechanics, DEM provides these high-resolution data by repeatedly applying Newton's equations of motion to each grain over time, after all applicable forces—primarily contact forces—are determined. Crucially, contact forces are usually calculated through a penalty-based formulation\footnote{There is another approach—the Non-smooth Contact Dynamics (NSCD)~\cite{jean1999non}—that enforces contact constraints strictly and finds contact impulses rather than forces via implicit instead of explicit solves. Just like any other numerical technique, both DEM and NSCD have their respective pros and cons; see~\cite{pazouki2017compliant} for a comparison and discussion in this regard.} that has three core parameters: the normal contact stiffness $k_n$ (which penalizes normal inter-grain overlap), the tangential contact stiffness $k_t$ (which penalizes tangential inter-grain slip), and the Coulomb-type contact friction $\mu_\text{p}$ (which bounds the frictional force's magnitude from above). 

Over the years, broadly speaking, DEM has evolved in two directions. The first direction stays with the common and classical ``rigid'' grains but extends DEM's modeling capability to complex grain shapes by clumping spheres~\cite{favier1999shape,li2022voxel} or using functions that include, but are not limited to, NURBS~\cite{andrade2012granular,lim2014granular}, Fourier descriptors~\cite{bowman2001particle,mollon20143d}, spherical harmonics~\cite{fan2024framework}, or level sets~\cite{kawamoto2016level,kawamoto2018all}. Along this direction, $k_n$ and $k_t$ are normally reviewed as numerical regularization parameters (although arguments exist in endowing them with physical interpretations~\cite{nadimi2022can,wang2025incorporating} from a model-fitting perspective), and $\mu_\text{p}$ is viewed as a material parameter; see also~\cite{odiscrete} for a related overview within the context of geomechanics. The second direction goes into relatively less common ``soft'' grains where grain deformation—induced by contact—becomes important and is resolved in a reduced-order manner; see~\cite{mollon2022soft,zunker2024mechanicallya,zunker2024mechanicallyb,zunker2025experimentally,henzel2026variational} for a few examples. In these studies, attention was mainly focused on frictionless contact-induced deformation, in which $k_n$ may be derived to depend on the local contact-overlap geometries and the constituent material's mechanical properties (e.g., elasticity). In the limit, one may go beyond DEM and into fully resolved approaches based on rigorous contact mechanics treatment~\cite{simo1992augmented,laursen2003computational} (despite their relatively much higher computational demand); see~\cite{liu2020ils,li2021emerging} as examples where the more challenging scenario of elastostatics was considered\footnote{NSCD has also been coupled with either MPM~\cite{nezamabadi2015implicit} or FEM~\cite{cardenas2020compaction} for modeling many contacting soft grains; these studies included inertia effects that make the problem less challenging to solve numerically 
than elastostatics scenarios.}. For these, $k_n$ (and $k_t$) recover their roles as numerical regularization parameters. Finally, a somewhat adjacent direction is more about pairing DEM with additional physics (and parallel approaches targeting soft contacting bodies, see for example~\cite{lara2025unified}); we do not expand our discussion in this regard, but three representative examples are adding grain bonding (e.g.,~\cite{potyondy2004bonded,harmon2021modeling}), background fluid (e.g.,~\cite{zhao2013coupled,wang2019novel}), and temperature (e.g.,~\cite{xiao2021dp,moncada2023level}).

Here, we propose a new framework to evolve DEM—also by \textit{shape} and along the first direction—but from a fundamentally distinct perspective that links a grain's surface topography (as the \textit{shape}) to $\mu_\text{p}$ (and $k_t$). Our motivation is primarily driven by the observation that current DEM frameworks—relying solely on contact friction—can fall short for dry granular systems composed of rough grains~\cite{pohlman2006surface,mollon2020can,pan2022influence}, highlighting the importance of grain surface roughness in certain scenarios. Adding another phenomenological parameter—for example, rolling friction~\cite{iwashita1998rolling}—could certainly offer some remediation~\cite{wilson2017influence,li2020identifying,pan2022influence}, but we are more interested in finding an alternative route that increases neither model complexity nor the level of phenomenology relative to current DEM frameworks. This route may, however, become (significantly) more computationally demanding, but we are less concerned given the rapid advances in computing capacity, especially when looking into the next ten or twenty years; in this context, see~\cite{negrut2012leveraging,zhang2024chrono} as representative examples and~\cite{zhou2026total1,zhou2026total2} as parallel examples for contacting deformable solids.

\begin{figure}
    \setlength{\abovecaptionskip}{0cm}
    \setlength{\belowcaptionskip}{0cm}
    \centering
    \includegraphics[width=0.85\linewidth]{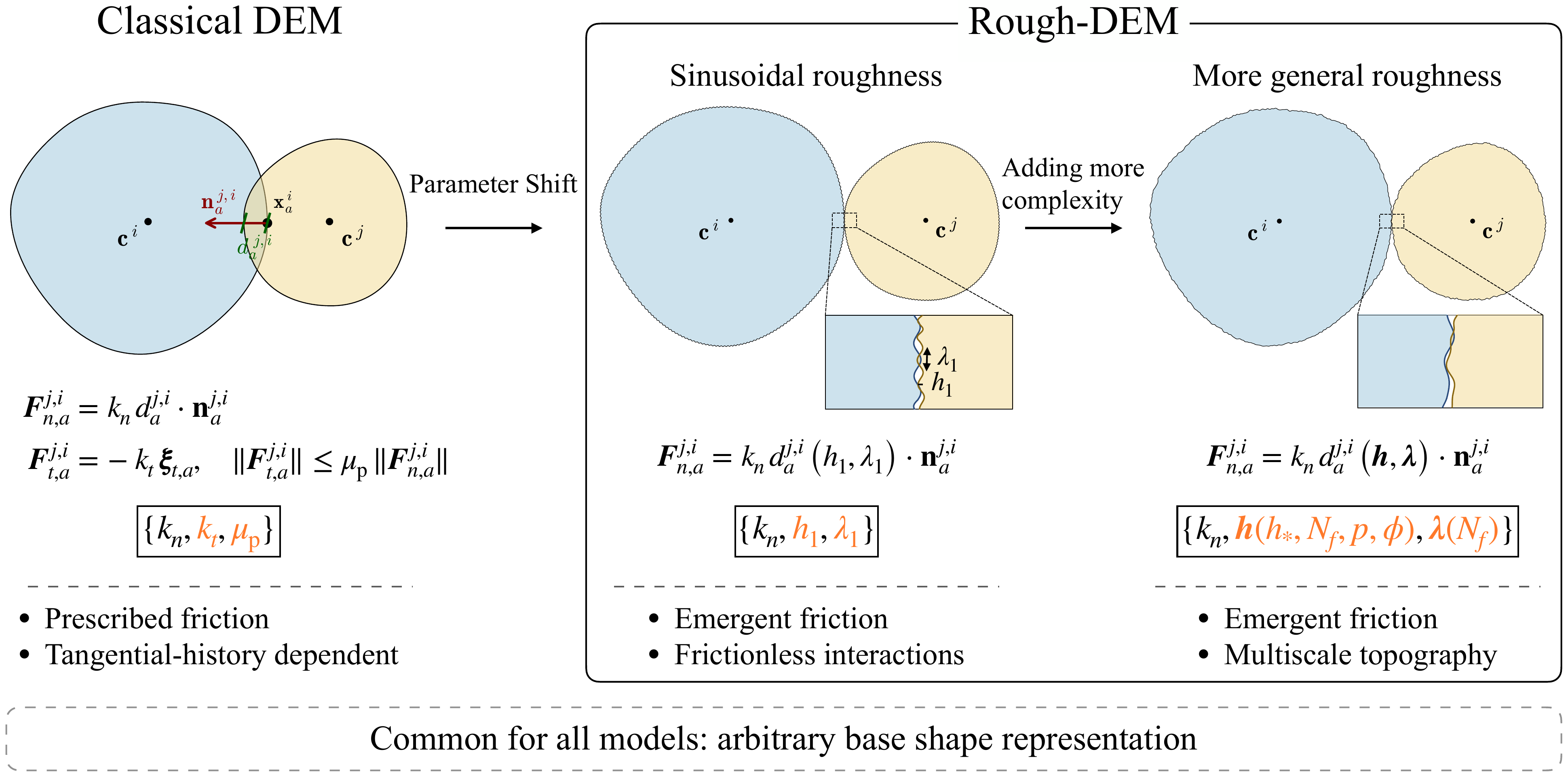}
    \caption{A 2D schematic illustration of the parameter shift from classical DEM with prescribed Coulomb-type friction to our Rough-DEM, in which contact friction emerges from frictionless interactions of surface asperities of contacting grains. For simple sinusoidal roughness, the amplitude $h_1$ and wavelength $\lambda_1$ are chosen as the dominant parameters. More complex surface roughness is represented using a Fourier series, with the (nominal) amplitude and wavelength vectors parameterized as $\boldsymbol{h}\left(h_\ast,N_f,p,\phi\right)$, $\boldsymbol{\lambda}(N_f)$, respectively; their dependence on $h_\ast,N_f,p$, and $\phi$ follows from the Fourier series formulation given in Eq.~\eqref{fourier_series_roughness}.}
    \label{fig:from_smooth_to_rough}
\end{figure}

So, more specifically, inspired by research from the tribology community, where the Amontons-Coulomb friction law (and many other variants) is reviewed as an emergent phenomenon from collective microscopic contacts~\cite{greenwood1966contact,bowden2001friction,sokoloff2001static}, we propose removing the need for a friction law and letting friction naturally emerge from collective frictionless interactions of surface asperities of contacting grains. From this perspective, herein we refer to our framework as roughness-resolved DEM, or in short, Rough-DEM. In practice, we propose a parameter shift from $\{k_n,k_t,\mu_\text{p} \}$—as is the case in all current DEM frameworks—to $\{k_n, \boldsymbol{h}, \boldsymbol{\lambda} \}$, where $\boldsymbol{h} = [h_1,h_2,...,h_k] \in \mathbb{R}^k$ and $\boldsymbol{\lambda} = [\lambda_1,\lambda_2,...\lambda_k]\in \mathbb{R}^k$ are geometric parameters by construction that describe, realistically when needed to be, the surface topography of a grain (details to be discussed later). Notably, in the special case where $k=1$ (i.e., $\boldsymbol{h} = h_1, \boldsymbol{\lambda}=\lambda_1$), our framework has the same number of model parameters as current frameworks, keeping the model complexity essentially unchanged; see Fig.~\ref{fig:from_smooth_to_rough} for an illustration\footnote{While modeling surface topography explicitly has been done, whether by intention or not (e.g., using spherical harmonics~\cite{fan2024framework} or clumping spheres~\cite{chen2024effects}, albeit the precision challenge associated with the latter), our framework fundamentally distinguishes from existing literature by eliminating the need for a friction law.}. In essence, we opt to trade phenomenology for physics at the cost of a likely increased computational demand. To mitigate the cost, we further introduce a remediation strategy that combines an FE$^2$-like algorithm, a neural-network surrogate, and a GPU-accelerated computation scheme (again, details to be discussed later). Further and more importantly, although the computational demand may still increase (modestly), the computing complexity in fact decreases because we no longer have to (1) keep storing and updating the tangential contact history, or (2) make an (unambiguous) decision on how to compute it (e.g., based on an overlapped distance-based approach~\cite{kawamoto2016level} versus on an overlapped volume-based one~\cite{feng2021energy,van2023physically}). Finally, it is worth noting that we do so on top of the Level Set-DEM~\cite{kawamoto2016level,kawamoto2018all}, such that Rough-DEM inherits the ability to model arbitrary grain shapes.

Our framework raises too many interesting questions to be addressed by this paper alone, but in this work, we show and discuss the following promising results obtained from Rough-DEM:
\begin{enumerate}
\item While friction depends on factors beyond just (usually complex) surface topography and is by itself a complex subject matter in tribology~\cite{chang1988static,pei2005finite,menezes2011role,albertini2021stochastic,bonari2022new,kuwik2023roughness, kalin2025understanding}, we show that a simple surface topography representation—parametrized by $h_1$ and $\lambda_1$—is sufficient to recover a two-body-based Amontons-Coulomb-type friction behavior within a DEM framework. By demonstrating its ability to recover this basic contact friction behavior, our framework opens the possibility of recovering more complex contact friction laws, when necessary, for predicting the macroscopic behaviors of granular materials.
\item While it remains unclear \textit{when and how exactly} grain surface roughness collectively influences the macroscopic behaviors of dry granular materials, we use the classical column collapse test to show that our framework, when parametrized by a simple surface topography construction (i.e., by $h_1$ and $\lambda_1$), can model scenarios within, and more importantly \textit{beyond}, the confines of current DEM frameworks under the same level of model complexity. We do so considering both circular- and noncircular-shaped grain assemblies, and we further demonstrate Rough-DEM's capability in modeling more complex surface topographies (parametrized by $\boldsymbol{h}$ and $\boldsymbol{\lambda}$). Our framework thus paves the way for investigating the collective effects of grain roughness in a more physics-grounded and systematic way, especially when paired with experimental investigations that exploit advanced manufacturing techniques.
\item While our framework is computationally more demanding by nature, we propose a remediation strategy that leverages
neural-network surrogates and GPU parallelization to accelerate computation. We are confident that computational demand will be increasingly less of a concern, given, again, rapid advances in computing software (including AI) and hardware capacities in the coming decades.
\end{enumerate}
Taken together, our framework, Rough-DEM, represents a paradigm shift in terms of how DEM is formulated for modeling (dry) granular material systems. Moreover, looking into granular-adjacent scientific domains, our framework could also be particularly useful for studying—and even designing—the rheology of dense frictional (non-brownian) suspensions composed of rough grains~\cite{hsiao2019experimental,singh2026jamming}, as well as the mechanics of contact-driven architected metamaterials~\cite{wang2021structured,lu2024role,zhou20253d}; for both scenarios, grain surface roughness—introduced inevitably by fabrication—can play an important role but its effects remain challenging to quantify systematically.

We organize the remainder of this paper as follows. Section.~\ref{sec:shape_representation_contact_computation} presents the techniques for representing the grain base geometry and surface topography, along with the corresponding contact detection and force computation procedures. Section.~\ref{sec:reproduce_friction} demonstrates Rough-DEM's compatibility with classical DEM and evaluates its ability to reproduce macroscopic frictional responses locally and collectively. Section.~\ref{sec:acceleration_strategy} introduces the main acceleration strategies for large-scale granular simulations. A FiLM-based neural network surrogate is also presented to accelerate signed distance evaluation. Finally, Section.~\ref{sec:conclusion_outlook} summarizes the main findings and discusses directions for future work.

\section{Roughness-resolved Grain Representation and Contact Computation}
\label{sec:shape_representation_contact_computation}

\subsection{Two-scale Geometric Representation}
\label{subsec:shape}
To represent grain geometry, we consider two distinct length scales: the base-shape scale and the roughness scale. The base shape describes the overall grain morphology, including its characteristic size, sphericity, and angularity, while the roughness component represents local topography, high-frequency deviations from the original smooth grain boundary/surface. Separating these two scales allows the higher-level grain morphology and the lower-level surface topography to be controlled and investigated independently. Moreover, it also avoids the need to increase complexity in the base representation in order to account for fine-scale asperity and interlocking effects.

First, the base shape of a grain is represented by a closed NURBS curve,
\begin{equation}\label{nurbs_expression}
    \vC(u) = \sum_{i=1}^nR_{i,p}(u)\vP_i,\quad u\in[0,1],
\end{equation}
where $R_{i,p}(u)$ denotes the NURBS basis functions, with $p$ being the degree of the basis function. $\vP_i$ is a vector comprising the coordinates of $n$ control points. The NURBS representation provides substantial flexibility for modifying the base shape by simply manipulating several control points.

The surface roughness is represented by a Fourier series defined on the parametric coordinate $u\in[0,1]$ of the base NURBS curve. Specifically, the roughness height is expressed as
\begin{equation}\label{fourier_series_roughness}
    h(u) = \sum_{k=1}^{N_f}\hat{h}_k\cos(2\pi ku+\phi_k),\quad \hat{h}_k = \frac{h_\ast}{k^p},
\end{equation}
where $\hat{h}_k$ and $\phi_k$ are the amplitude and phase of the $k$-th Fourier mode. $h_\ast$ is the roughness amplitude parameter, and $p$ is the spectral decay exponent that controls the rate at which Fourier-mode amplitudes decrease with increasing wavenumber. $N_f$ denotes the number of Fourier modes. The Fourier representation provides a flexible multiscale description of surface roughness and can approximate complex or statistically irregular surface profiles with increasing accuracy as additional modes are retained. In particular, $N_f$ determines the finest roughness scale represented in the parametric domain, thereby providing a direct means of balancing geometric fidelity against computational cost.

Combining Eqs.~\eqref{nurbs_expression} and \eqref{fourier_series_roughness}, the final rough grain boundary $\tilde{\vC}^i$ of the $i$-th grain is constructed by offsetting the base curve along its local outward normal $\vn^i$, i.e.,
\begin{equation}\label{roughness_description}
    \tilde{\vC}^i(u) = \vC^i(u)+h^i(u)\cdot\vn^i(u),
\end{equation}
where $\vn^i$ denotes the outward unit normal vector to the base curve $\vC^i$ and is computed from the first derivative of the NURBS representation.

\subsection{Contact Detection}
\label{subsec:contact_detection}
Penalty-based contact computation relies on the determination of penetration depth or overlap area between contacting bodies, which in turn necessitates the continuous evaluation of signed distances at potentially contacting points in penetration-based formulations and the measurement of the overlapping region in area-based formulations. For grains with fine-scale, rapidly varying surface roughness, however, such geometric evaluations can be computationally expensive. To address this challenge, we adopt a FE$^{2}$-like strategy that leverages the two-scale shape representation introduced in Section~\ref{subsec:shape}: coarse screening at the base-shape scale, followed by roughness-resolved contact detection within the identified candidate regions. 

\begin{figure}[!htbp]
    \setlength{\abovecaptionskip}{0cm}
    \setlength{\belowcaptionskip}{0cm}
    \centering
    \includegraphics[width=0.85\linewidth]{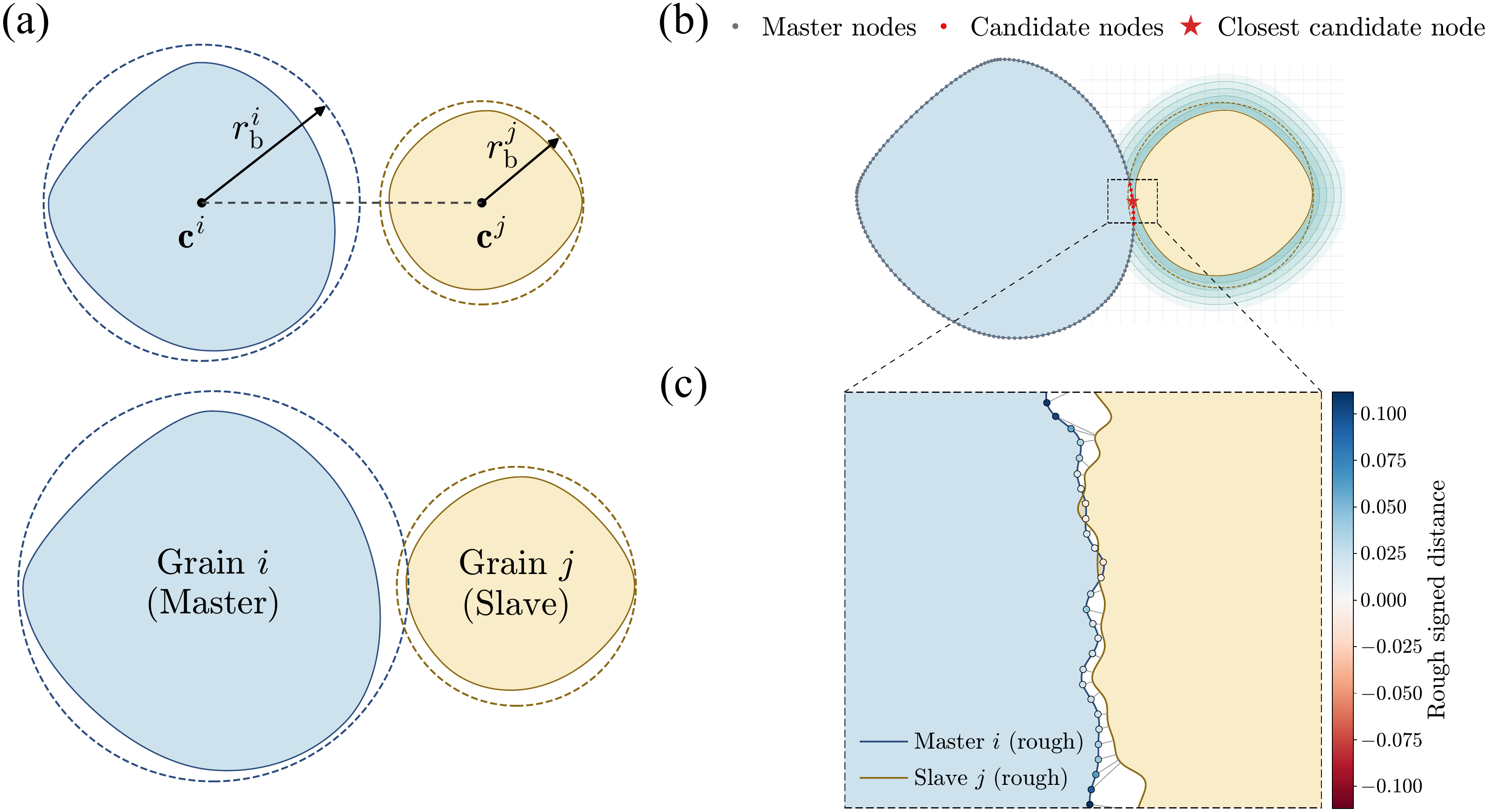}
    \caption{Hierarchical contact detection procedure for rough grains: (a) coarse contact screening using bounding circles; (b) finer contact screening using an LS-DEM-like contact evaluation~\cite{kawamoto2016level} to identify candidate master nodes within the slave bounding circle and then construct a local window centered on the nearest candidate node; and (c) finest and final contact screening and computation by evaluating roughness-scale signed distances within the local window, where negative and positive values indicate penetration and separation, respectively.}
    \label{fig:hierarchical_contact}
\end{figure}

At each time step, contact detection is carried out hierarchically, as illustrated in Fig.~\ref{fig:hierarchical_contact}. The coarse-screening stage is intended to eliminate clearly inactive grain pairs and to identify potentially contacting boundary regions, without explicitly constructing the fine-scale roughness. Specifically, each grain is enclosed by a circular bounding region centered at its centroid, with a radius that covers the entire base geometry plus an additional margin to account for the amplitude of roughness perturbation. A grain pair is retained only if
\begin{equation}
    \|\vc^i-\vc^j\| \leq r_\rb^i+r_\rb^j,
\end{equation}
where $\vc^i$ and $r_\rb^i$ denoting the centroid and bounding radius of grain $i$, respectively. This provides a relatively inexpensive screening criterion that requires only grain centroids and precomputed bounding radii.

For each retained pair, the coarse boundary nodes of the master grain are further examined. Only those located within the bounding region of the slave grain are kept as candidate nodes. If no candidate node is identified, the grain pair is excluded from subsequent rough-level evaluation. Otherwise, these candidate nodes are transformed into the body-fixed frame of the slave grain, where their signed distances to the slave base geometry are efficiently evaluated using a level-set parametrization, as in~\cite{kawamoto2016level}. The node with the minimum coarse-scale signed distance is then selected as the center of the potentially contacting region, as indicated by the star in Fig.~\ref{fig:hierarchical_contact}(b).

Within the identified local window, the detailed surface roughness is further resolved for fine-scale contact detection. The refined master nodes are then passed to the rough-level detector, where signed distances and surface normals relative to the rough geometry of the slave grain are computed using closest-point projection \cite{konyukhov2008solvability}. This two-stage procedure confines the computationally expensive rough-scale geometric queries to a small portion of the grain boundary selected through inexpensive coarse screening. A level-set-based approach is not adopted here at the roughness scale because of the significant separation between the roughness and grain length scales, which would require an excessively dense grid to resolve the fine surface features accurately.

\subsection{Grain Kinematics and State Update}
\label{subsec:kinematics_state_update}
\begin{figure}[!htbp]
    \setlength{\abovecaptionskip}{0cm}
    \setlength{\belowcaptionskip}{0cm}
    \centering
    \includegraphics[width=0.5\linewidth]{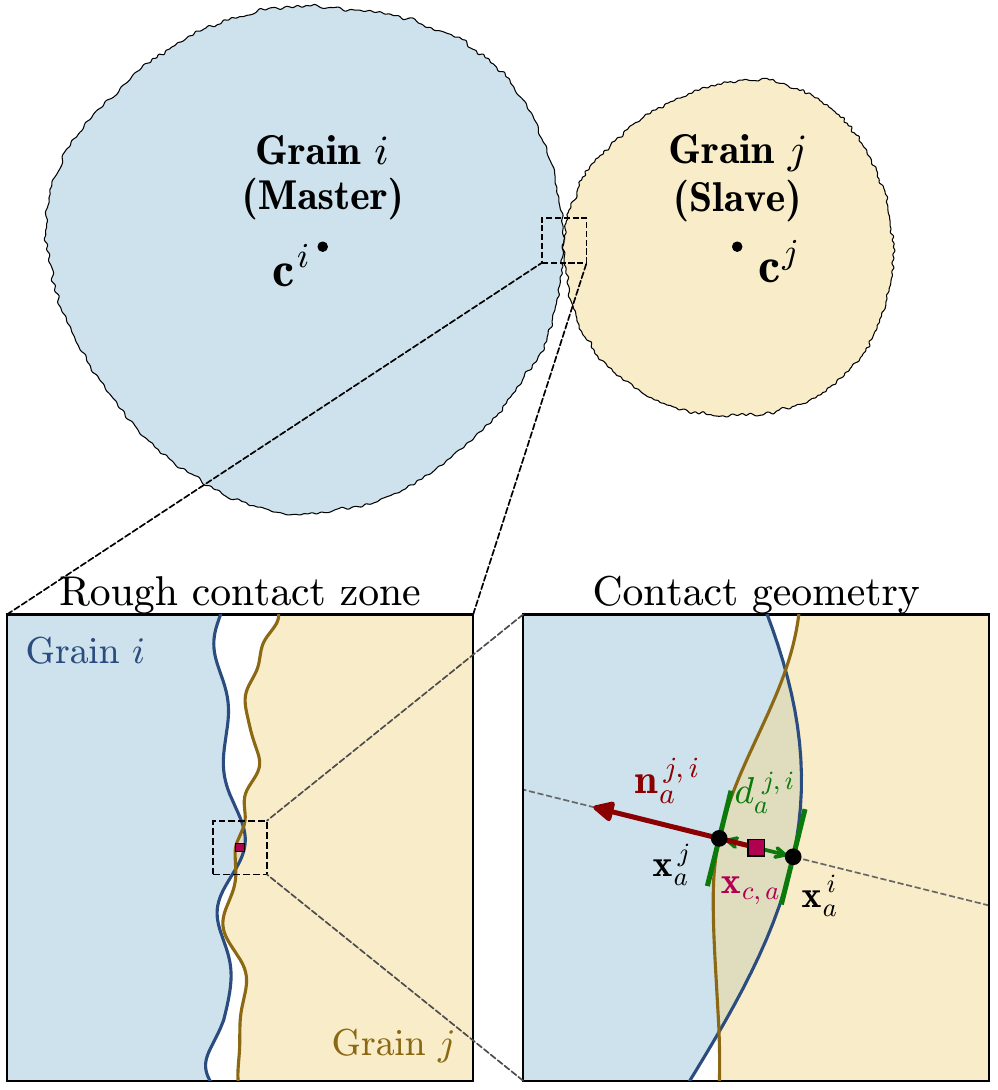}
    \caption{Schematic of the rough-contact geometry between two grains. Grain $i$ is designated as the master and grain $j$ as the slave, with centroids $\vc^i$ and $\vc^j$. For the $a$-th penetrated master node, $\vx_a^i$ is the node position and $\vx_a^j$ is its projection onto the slave surface along the outward unit normal $\vn_a^{j,i}$. The signed distance $d_a^{j,i}$ is negative under penetration, and $\vx_{c,a} = \left(\vx_a^i+\vx_a^j\right)/2$ defines the common contact point. The subscripts a and c denote the master-node index and ``common'', respectively. A single superscript indicates grain ownership, whereas the ordered superscript pair $(j,i)$ identifies a quantity defined for master grain $i$ relative to slave grain $j$; for the contact normal, it also specifies the direction from $j$ to $i$.}
    \label{fig:contact_diagram_rough}
\end{figure}

With the identification of active roughness-scale contact nodes, the kinematics of contacting grains can be subsequently determined. As shown in Fig.~\ref{fig:contact_diagram_rough}, for each penetrated master node $\vx_a^i$, the corresponding projection point on the slave surface is obtained by projecting along the outward normal $n_a^{j,i}$ directed from grain $j$ toward grain $i$, i.e., $\vx_a^j = \vx_a^i-d_a^{j,i}\vn_a^{j,i}$. The common contact point of the two grains is then defined as the midpoint between the penetrating master node $\vx_a^i$ and its projection point $\vx_a^j$ on the slave surface,
\begin{equation}
    \vx_{c,a} = \frac{1}{2}\left(\vx_a^i+\vx_a^j\right) = \vx_a^i-\frac{1}{2}d_a^{j,i}\vn_a^{j,i},
\end{equation}
where $d_a^{j,i}\leq0$ is the rough signed distance of the $a$-th nodal seed, $a\in\{a\in\mathbb{Z}\,|\,1\leq a\leq A\}$ with $A$ being the number of nodes used to discretize the rough grain $i$. The projected point is used as the representative contact location for both grains.

A linear spring–dashpot contact model is adopted to relate the local penetration depth to the contact force. Let $\Gamma_c^i$ denote the active portion of the boundary of grain $i$ that penetrates grain $j$. The resultant normal contact force acting on grain $i$ is obtained by integrating the normal contact traction $\vt_n^i(s)$ along this boundary, which can be further approximated by summation over all refined nodes, i.e.,
\begin{equation}\label{total_normal_force}
    \vF_{n}^i = \int_{\Gamma_c^i}\vt_n^i(s)\,\rd s \approx \sum_{a=1}^A\vt_{n,a}^i\Delta s_a,
\end{equation}
where $\Delta s_a$ is the arc-length weight associated with node $\vx_a^i$. For an interior node, this weight is evaluated from the two adjacent boundary segments as
\begin{equation}\label{arc_length_weight}
    \Delta s_a = \frac{1}{2}\left(\|\vx_a^i-\vx_{a-1}^i\|+\|\vx_{a+1}^i-\vx_a^i\|\right),
\end{equation}
while one-half of the adjacent segment length is assigned to each endpoint of the penetrated roughness curve. This provides a mesh-independent approach (similar to~\cite{feldfogel2024discretization}) for computing the contact force by weighting each nodal contribution according to its associated curve length, thereby minimizing sensitivity to the spatial discretization~\cite{li2019capturing} of the surface roughness.

The nodal evaluation of the distributed traction in Eq.~\eqref{total_normal_force} takes the form
\begin{equation}\label{distributed_traction_nodal}
    \vt_{n,a}^i = \frac{1}{L_\mathrm{ref}^i}\left[k_n\delta_{n,a}-\gamma_nv_{n,a}\right]_+\vn_a^{j,i},
\end{equation}
where $L_\mathrm{ref}^i$ represents the reference perimeter of grain $i$, introduced as the characteristic length to normalize the arc-length contribution; $k_n$ is the normal contact stiffness; and $\gamma_n$ is the normal damping coefficient, which is typically determined from the prescribed normal coefficient of restitution $e_n$ in a binary collision \cite{da2005rheophysics}. Here, $[x]_+=\max(x,0)$ denotes the positive-part operator. It enforces a non-cohesive contact response, so that the normal interaction cannot generate attractive forces. The local penetration depth is obtained directly from the rough signed distance of the corresponding node through
\begin{equation}\label{penetration_depth}
    \delta_{n,a} = \max\left[-d_a^{j,i},0\right],
\end{equation}
while the normal velocity term $v_{n,a}$ in Eq.~\eqref{distributed_traction_nodal} is defined as the signed normal component of the relative velocity at $\vx_c(s)$,
\begin{equation}\label{relative_normal_velocity}
    v_{n,a} = \left(\vv_{c,a}^i-\vv_{c,a}^j\right)\cdot\vn_a^{j,i},
\end{equation}
such that $v_{n,a}<0$ corresponds to approaching motion and $v_{n,a}>0$ to separating motion. With this convention, the damping term $-\gamma_n v_{n,a}$ always opposes the normal relative motion.

The contact-point velocity $\vv_{c,a}^i$ follows the rigid-body motion of two grains,
\begin{equation}
    \vv_{c,a}^i = \vv^i + \boldsymbol{\omega}^i\times\left(\vx_{c,a}-\vc^i\right),
\end{equation}
where $\vv^i$ and $\boldsymbol{\omega}^i$ are the translational and angular velocities of grain $i$, respectively, and are updated according to Eqs.~\eqref{translational_velocity_update} and \eqref{rotational_velocity_update}.

By Newton's third law, the force contribution applied to grain $j$ should be equal and opposite, i.e., $\vF_n^j = -\vF_n^i$. The contact moment magnitude about the centroids of the two grains is assembled from the same nodal forces as
\begin{equation}\label{total_normal_moment}
    M_n^i = \sum_{a=1}^A\left(\vx_{c,a}-\vc^i\right)\times\vF_{n,a}^i\cdot\ve_3.
\end{equation}

After all contact contributions have been assembled over all active interactions, the translational and angular accelerations of grain $i$ are obtained from the Newton–Euler equations,
\begin{equation}
    \va_\ell^i = \frac{\vF_\ell^i}{m^i},\quad \alpha_\ell^i = \frac{M_\ell^i}{I^i},
\end{equation}
where the subscript $\ell$ denotes the discrete time-step index, $t_\ell = \ell\Delta t$, and $\vF_\ell^i$ and $M_\ell^i$ are the assembled force and moment applied on grain $i$ at that time. Here $m^i=\rho hA^i$ and $I^i=\rho hJ^i$, where $h=1\,\mathrm{mm}$ is the assumed out-of-plane thickness, and $A^i$ and $J^i$ denote the grain area and polar area moment, respectively. Since our focus is the frictionless interaction of roughness, we should have $\vF_\ell^i=\vF_{n,\ell}^i,\ M_\ell^i=M_{n,\ell}^i$. The grain state is then advanced using an explicit central difference scheme, in which the linear viscous damping term is evaluated using a centered approximation:
\begin{subequations}\label{velocity_update}
    \begin{align}
        \vv_{\ell+1}^i &= \frac{\left(1-\frac{\eta\Delta t}{2}\right)\vv_\ell^i+\va_\ell^i\Delta t}{1+\frac{\eta\Delta t}{2}}; \label{translational_velocity_update}\\
        \omega_{\ell+1}^i &= \frac{\left(1-\frac{\eta\Delta t}{2}\right)\omega_\ell^i+\alpha_\ell^i\Delta t}{1+\frac{\eta\Delta t}{2}}, \label{rotational_velocity_update}
    \end{align} 
\end{subequations}
where $\eta$ is the global damping coefficient. The centroid position and orientation are subsequently updated as
\begin{equation}\label{position_orientation_update}
    \vc_{\ell+1}^i = \vc_\ell^i + \vv_{\ell+1}^i\Delta t,\quad \theta_{\ell+1}^i = \theta_\ell^i + \omega_{\ell+1}^i\Delta t.
\end{equation}

Finally, the boundary nodes are updated through the corresponding rigid-body translation and rotation, that is,
\begin{equation}\label{rigid_body_transformation}
    \vx_{a,\ell+1}^i = \vc_{\ell+1}^i + \vR(\Delta\theta_\ell^i)\left(\vx_{a,\ell}^i-\vc_\ell^i\right),\quad \Delta\theta_\ell^i = \omega_{\ell+1}^i\Delta t,
\end{equation}
where $\vR$ is the two-dimensional rotation matrix associated with the incremental rotation $\Delta\theta_\ell^i$.

\section{Roughness-induced Macroscopic Friction}
\label{sec:reproduce_friction}

\subsection{Reduction to Classical Frictionless DEM}
\label{subsec:compatibility_with_dem}
Before investigating the friction-like effects generated by explicitly resolved surface roughness, it is necessary to verify that the proposed Rough-DEM recovers the classical DEM in the smooth-grain limit. This comparison provides a baseline verification of the contact detection and time integration procedures, ensuring that the friction-like resistance observed in rough grains originates from surface geometry and the resulting evolution of local contact normals, rather than numerical artifacts or an explicitly prescribed tangential friction law.

\begin{figure}[!htbp]
    \setlength{\abovecaptionskip}{0cm}
    \setlength{\belowcaptionskip}{0cm}
    \centering
    \includegraphics[width=0.3\linewidth]{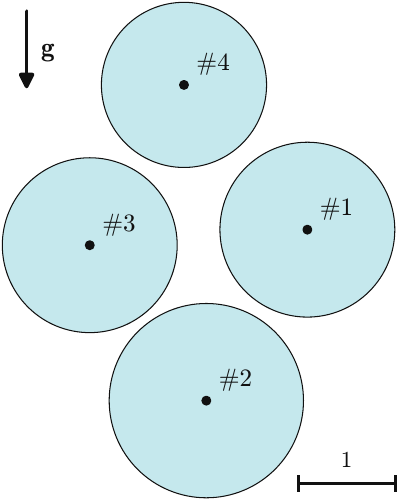}
    \caption{Initial configuration of the four-grain system. Grains $\#1,\#3$, and $\#4$ are released from rest under gravitational acceleration $\vg$, while grain $\#2$ is held fixed and serves as an obstacle.}
    \label{fig:initial_grain_position}
\end{figure}

To this end, we consider a four-grain system subjected to gravity, as shown in Fig.~\ref{fig:initial_grain_position}. The grain radii are $R=[0.9, 1.0, 0.9, 0.85]\,\mathrm{mm}$, with a density of $\rho=2.4\times10^{-6}\,\mathrm{kg/mm^3}$. All grain surfaces are rendered smooth by setting the roughness amplitude $h_\ast=0$, while the friction coefficient in the reference DEM is set to $\mu=0$. Both simulations employ identical contact and numerical parameters, including a normal contact stiffness of $k_n=2353.608\,\mathrm{N/m}$, a coefficient of restitution of $c_n=1.0$, and a global damping coefficient of $\eta=0.9\,\mathrm{s^{-1}}$. The gravitational acceleration is $g=-9806.7\,\mathrm{mm/s^2}$, and a time step of $\Delta t=10^{-5}\,\mathrm{s}$ is used for a total of 2000 simulation frames. All other initial conditions and numerical settings are kept identical between the two simulations.

\begin{figure}[!htbp]
    \setlength{\abovecaptionskip}{0cm}
    \setlength{\belowcaptionskip}{0cm}
    \centering
    \subfigure[$x$-trajectory]{
        \includegraphics[width=0.475\linewidth]{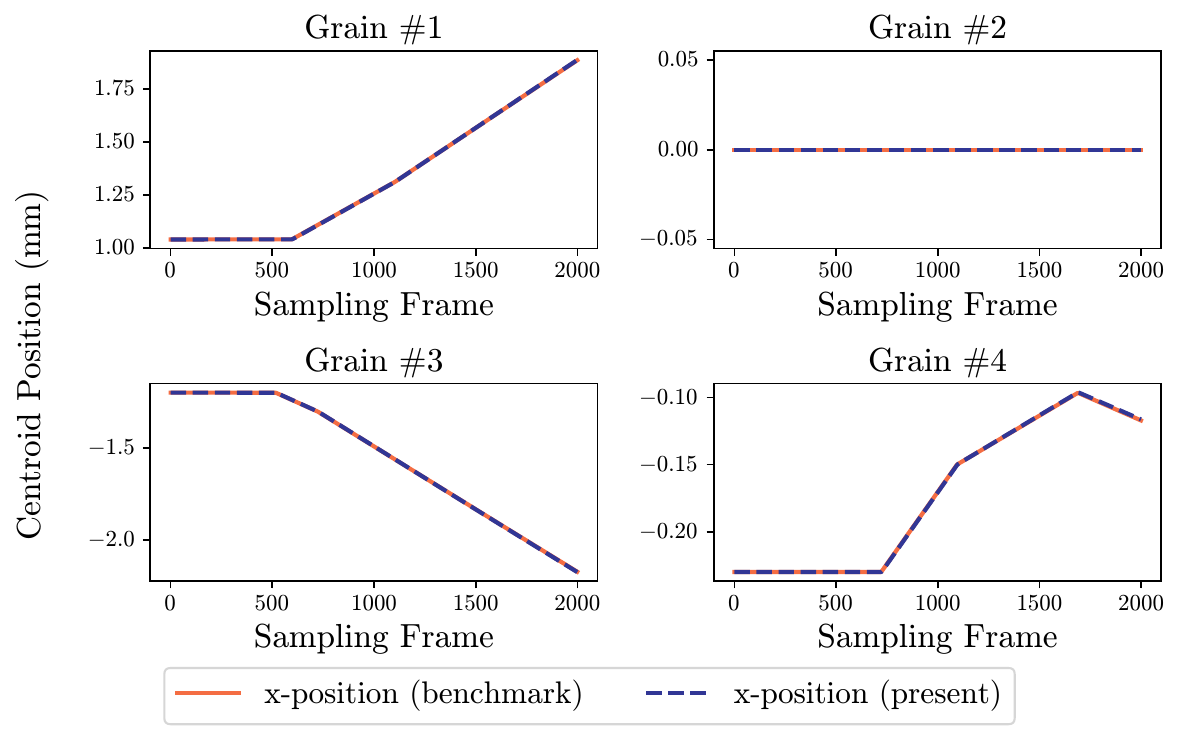}
    }
    \subfigure[$y$-trajectory]{
        \includegraphics[width=0.475\linewidth]{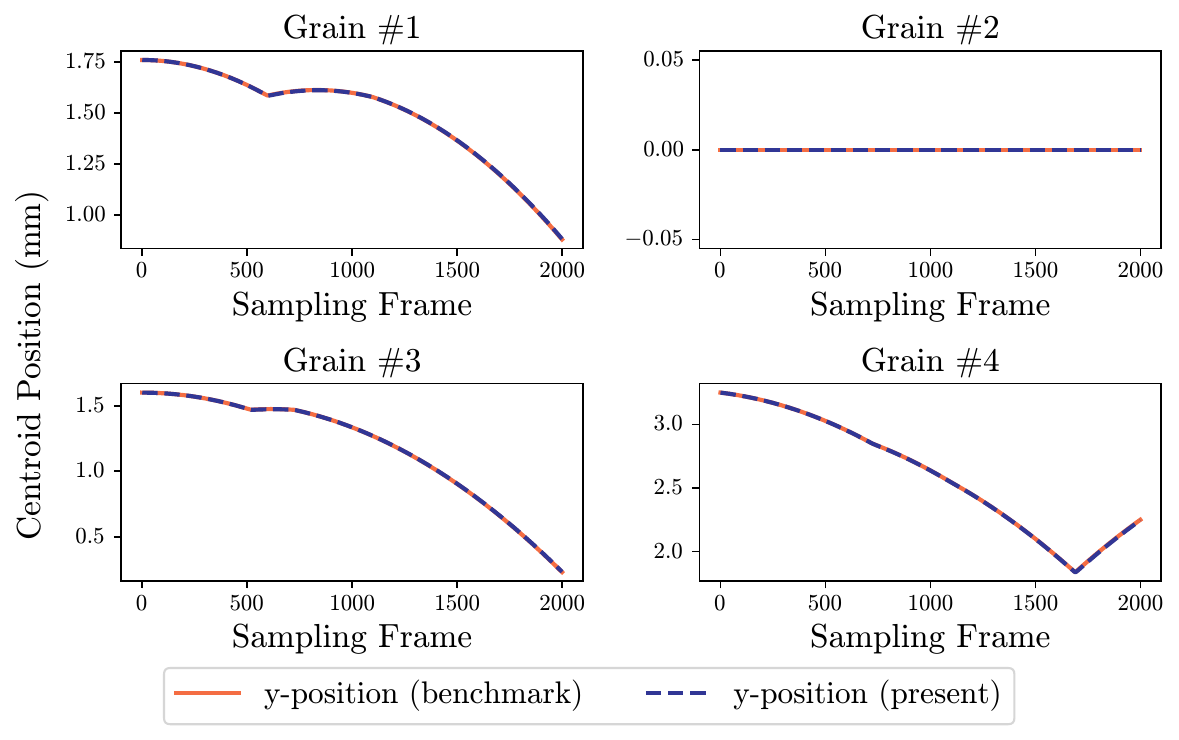}
    }
    \subfigure[$x$-velocity]{
        \includegraphics[width=0.475\linewidth]{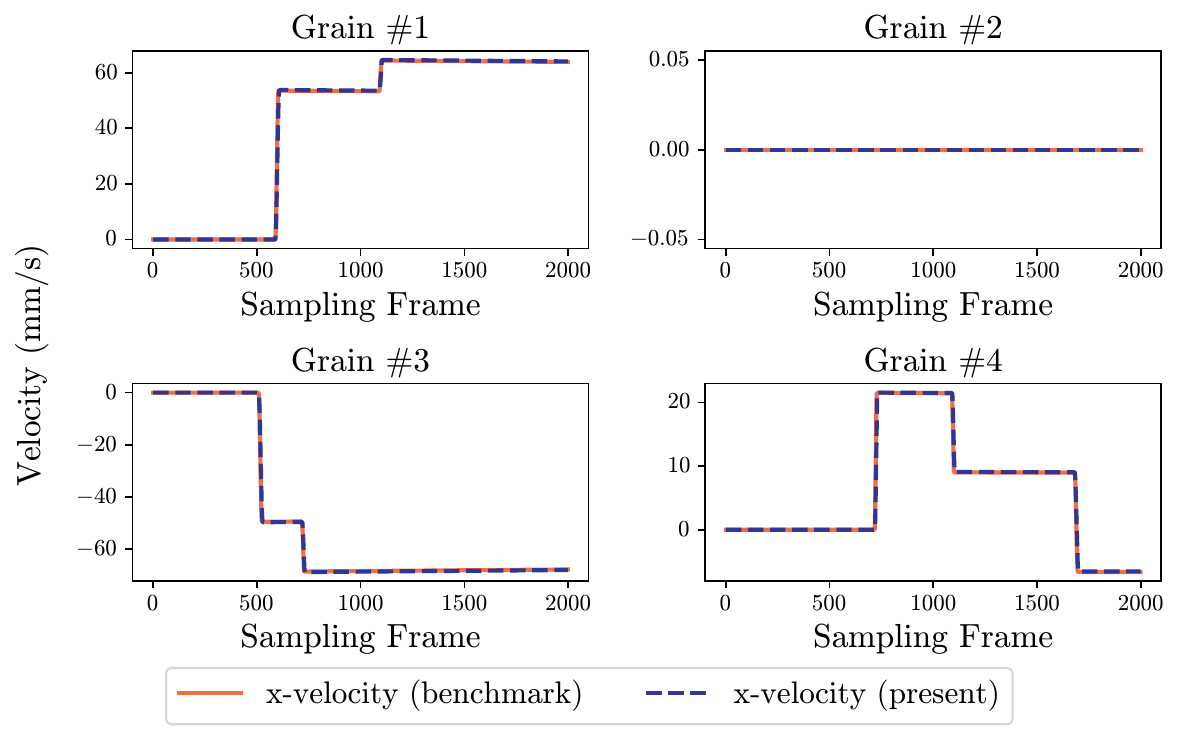}
    }
    \subfigure[$y$-velocity]{
        \includegraphics[width=0.475\linewidth]{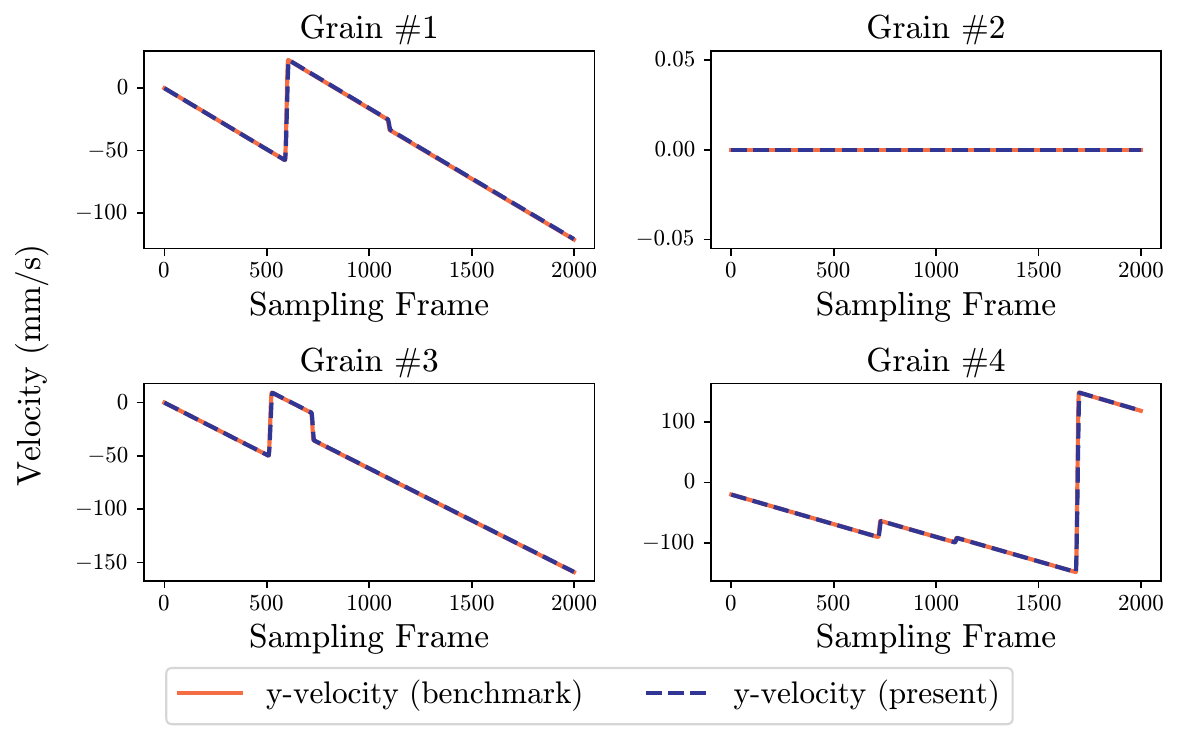}
    }
    \caption{Comparison of the trajectories and velocity histories given by the Rough-DEM and the reference DEM for the four-grain system shown in Fig.~\ref{fig:initial_grain_position}.}
    \label{fig:dem_comparison}
\end{figure}

As shown in Fig.~\ref{fig:dem_comparison}, the two models exhibit excellent agreement in both grain centroid trajectories and velocity evolution, confirming that the proposed Rough-DEM successfully recovers the classical frictionless DEM response in the smooth-grain limit.

\subsection{Recovery of Local Frictional Behavior: Sliding Block Test}
\label{subsec:recover_local_friction}
\begin{figure}[!htbp]
    \setlength{\abovecaptionskip}{0cm}
    \setlength{\belowcaptionskip}{0cm}
    \centering
    \includegraphics[width=0.42\linewidth]{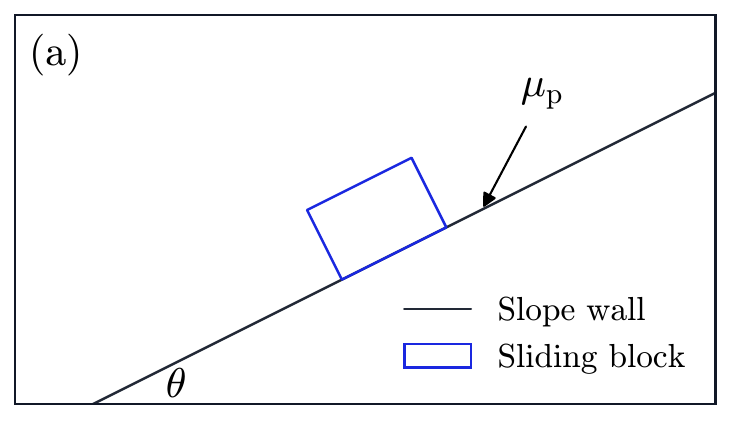}
    \includegraphics[width=0.42\linewidth]{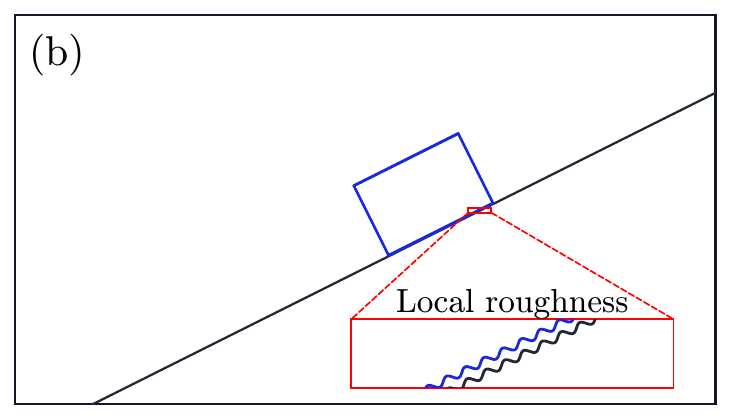}
    \caption{Schematic of the sliding block test: (a) classical DEM configuration with a Coulomb friction law; (b) Rough-DEM configuration incorporating explicit local (sinusoidal) roughness structures.}
    \label{fig:sliding_block_schematic}
\end{figure}

We now investigate whether explicitly resolved surface asperities can reproduce the macroscopic frictional behavior observed in classical DEM simulations. Friction-like responses are assessed at both the local and collective scales to evaluate the effectiveness of the Rough-DEM framework and to determine whether deviations in local frictional behavior translate into appreciable effects at the collective scale. In this subsection, the recovery of local frictional behavior is assessed using the well-established sliding-block test shown in Fig.~\ref{fig:sliding_block_schematic}.

In the context of the classical Coulomb friction law, the tangential acceleration of the block is governed by the competition between the downslope component of gravity and the frictional resistance. Once sliding occurs, the analytical tangential acceleration is
\begin{equation}\label{analytical_acceleration}
    a_t = -g\sin\theta+\mu_\text{p}\, g\cos\theta,
\end{equation}
where $\mu_\text{p}$ is the prescribed friction coefficient in the Coulomb friction model. In the roughness-resolved model, no explicit tangential force or friction model is prescribed. Instead, resistance to sliding emerges solely from frictionless normal interactions between the resolved surface asperities. The sliding block test is therefore used to examine whether these microscopic interactions can give rise to a finite macroscopic resistance to tangential motion and, if so, whether the resulting response can be interpreted in terms of an effective friction coefficient.

For simplicity, the contacting surfaces are represented by sinusoidal roughness, which can be regarded as a special case of the Fourier representation introduced in Eq.~\eqref{fourier_series_roughness}. Specifically, $N_f=1$, so that $\hat{h}_1=h_\ast$, while the decay exponent \(p\) is irrelevant for the present single-mode case. Introducing the normalized local coordinate $u=s/\lambda$, where $s$ denotes the arc-length coordinate along the mean contact surface, the roughness profile is written as
\begin{equation}\label{sinusoidal_roughness_profile}
    h(s)=h_\ast\cos\left(2\pi s/\lambda+\phi_1\right),
\end{equation}
with $\hat{h}_1=h_\ast=1.6\times 10^{-3}\,\mathrm{mm},\ \lambda=0.025\,\mathrm{mm},$ and $\phi_1=-\frac{\pi}{2}$, giving the equivalent sinusoidal form $h(s) = h_\ast\sin\left(2\pi s/\lambda\right)$. The corresponding spatial frequency is $1/\lambda=40\,\mathrm{mm^{-1}}$. The same roughness amplitude and wavelength are assigned to the block and the inclined surface. The rectangular block has dimensions $1.5\,\mathrm{mm}\times 1.0\,\mathrm{mm}$, so that 60 roughness periods are distributed along its lower surface. The inclined plane is defined by $y=kx$ over $0\leq x\leq50\,\mathrm{mm}$, giving a physical length $L_\mathrm{slope}=50\sqrt{1+k^2}\,\mathrm{mm}$.

\begin{figure}[!htbp]
    \setlength{\abovecaptionskip}{0cm}
    \setlength{\belowcaptionskip}{0cm}
    \centering
    \includegraphics[width=0.49\linewidth]{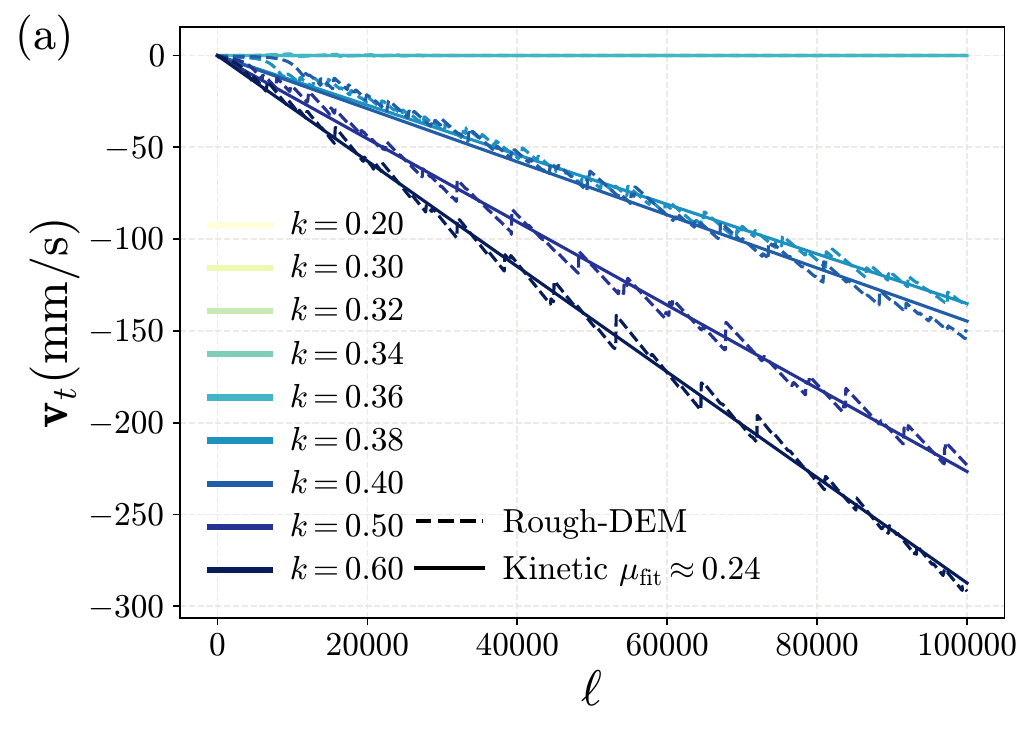}
    \hfill
    \includegraphics[width=0.49\linewidth]{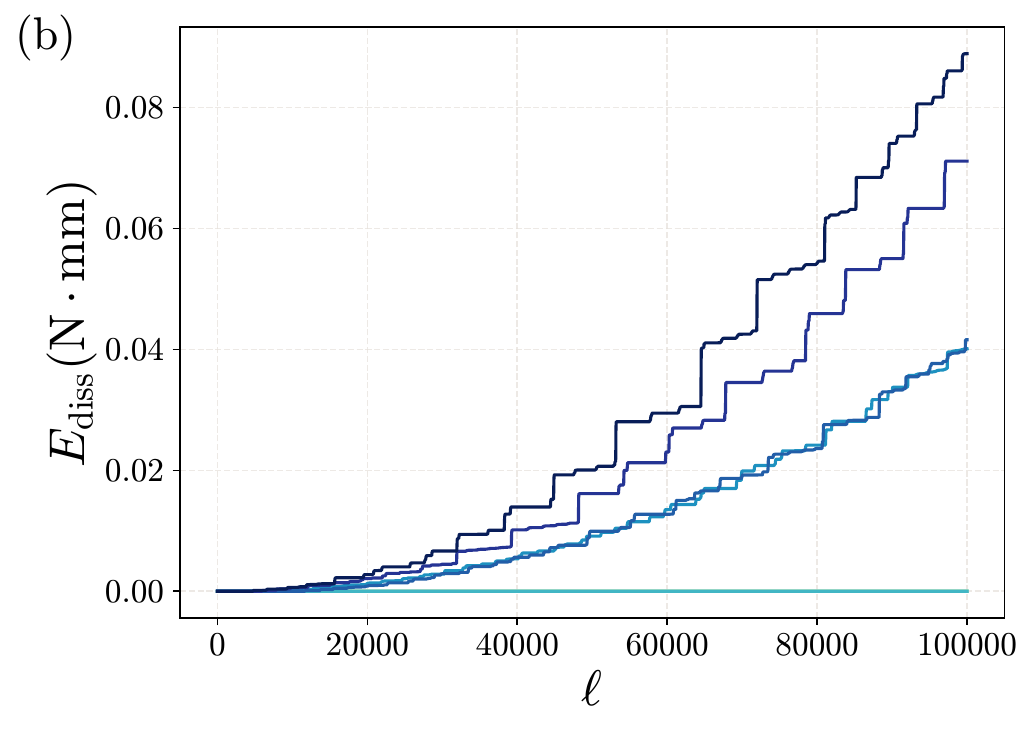}
    \caption{Sliding block test using our Rough-DEM: (a) tangential velocities of the block at various slope angles. Dashed lines represent the simulation results, while solid lines denote least-squares fits for the sliding cases, corresponding to an effective macroscopic friction coefficient of approximately 0.24; (b) cumulative energy dissipated through normal contact interactions over the course of the simulation. Recall that $\ell$ denotes the discrete time-step index, with $t_\ell=\ell\Delta t$.}
    \label{fig:tangential_velocity_dissipation}
\end{figure}

To minimize the sensitivity to the initial asperity phase, the block is initially positioned near $x=30\,\mathrm{mm}$, with its exact horizontal location $x_\mathrm{devi}$ slightly adjusted so that the roughness profiles of two contacting surfaces are phase-aligned across various slopes $k$. A small initial offset normal to the slope, $\epsilon_n = 0.1h_\ast$, is prescribed to reduce the potential influence of the initial falling motion.

Contact between the rough surfaces is modeled using a normal spring-dissipation law, with no explicit tangential contact force. The normal stiffness is set to $k_n=1\times 10^4\,\mathrm{N/m}$, the material density to $\rho=2.4\times 10^{-6}\,\mathrm{kg/mm^3}$, and the gravitational acceleration to $g=9806.7\,\mathrm{mm/s^2}$. The time step here is $\Delta t=1\times 10^{-6}\,\mathrm{s}$, and each simulation is advanced for $10^5$ steps, corresponding to a total physical duration of $0.1\,\mathrm{s}$. The normal coefficient of restitution $e_n$ is set to zero, while the global damping coefficient $\eta$ is also set to zero, so that dissipation arises only through the local normal contact interaction rather than through an artificial system-wide damping term.

We record the tangential velocities of the block for $k=$ 0.2, 0.3, 0.32, 0.34, 0.36, 0.38, 0.4, 0.5, and 0.6, as shown in Fig.~\ref{fig:tangential_velocity_dissipation}(a). The results clearly capture both the stick-to-slide transition and tangential resistance, two characteristic behaviors commonly observed in frictional systems. The transition occurs between $k=0.36$ and $k=0.38$, indicating an effective static friction coefficient of $\mu_\mathrm{s}\in[0.36,0.38)$. Least-squares fits to the sliding cases yield approximately the same slope, from which a uniform macroscopic friction coefficient, $\mu_\mathrm{fit}\approx 0.24$, can be inferred and interpreted as the effective kinetic friction coefficient. The resulting separation between $\mu_\mathrm{s}$ and $\mu_\mathrm{k}$, with $\mu_\mathrm{s}>\mu_\mathrm{k}$, is consistent with the commonly observed distinction between static and kinetic friction \cite{rabinowicz1951nature,persson2003nature}.

The roughness-induced resistance is also accompanied by dissipation through the normal contact interaction (Eq.~\eqref{distributed_traction_nodal}). The total instantaneous dissipation rate is obtained by summing the dissipative term over all active contact nodes,
\begin{equation}
    P_\mathrm{diss}(t) = \sum_{a\in\mathcal{C}}\frac{\gamma_n}{L_\mathrm{ref}^i}v_{n,a}^2\Delta s_a,
\end{equation}
where $\mathcal{C}$ denotes the set of active non-cohesive contacts. Since the damping force always opposes the signed normal relative motion, $P_{\mathrm{diss}}\geq0$ represents the rate at which mechanical energy is irreversibly removed from the system. The cumulative energy dissipated through normal contact is then
\begin{equation}\label{normal_energy_dissipation}
    E_\mathrm{diss}(t) = \int_0^t P_\mathrm{diss}(\tau)\,\rd\tau.
\end{equation}

Fig.~\ref{fig:tangential_velocity_dissipation}(b) shows that $E_\mathrm{diss}$ remains negligible for the sticking cases but increases progressively once sustained sliding develops. Its step-like evolution reflects the intermittent nature of roughness-scale interactions: dissipation is concentrated during individual loading and collision events rather than being continuously imposed through a prescribed friction law. Physically, the rough surface continuously redirects part of the macroscopic tangential motion into local normal motion, which is subsequently dissipated through the normal damping mechanism.

\begin{figure}[!htbp]
    \setlength{\abovecaptionskip}{0cm}
    \setlength{\belowcaptionskip}{0cm}
    \centering
    \includegraphics[width=0.49\linewidth]{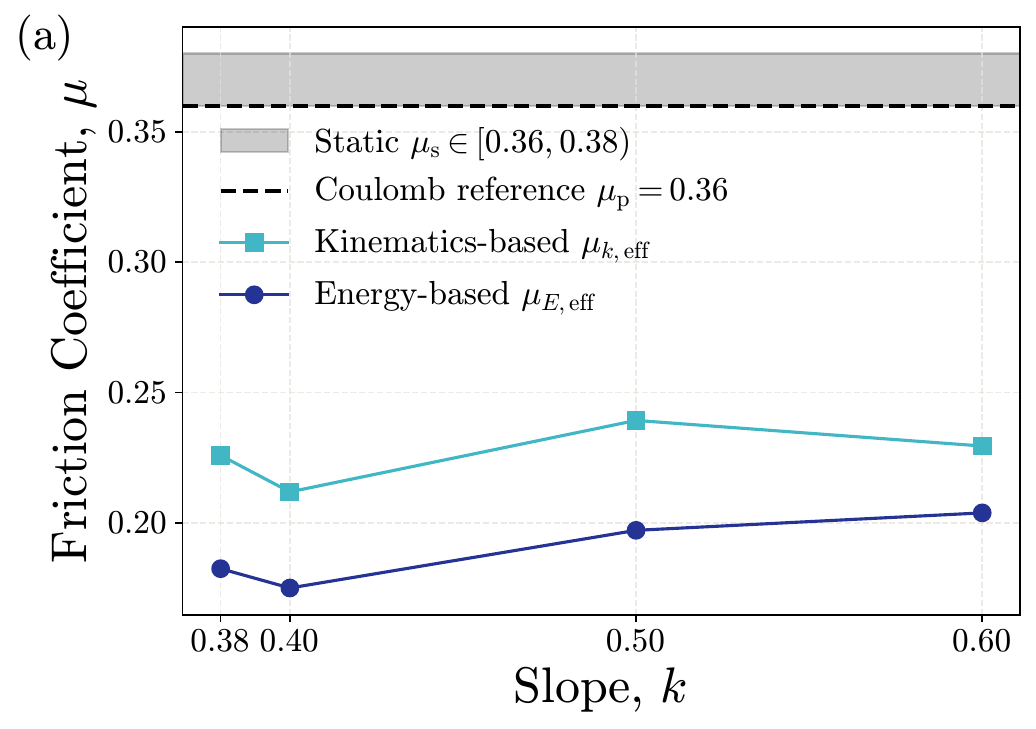}
    \hfill
    \includegraphics[width=0.49\linewidth]{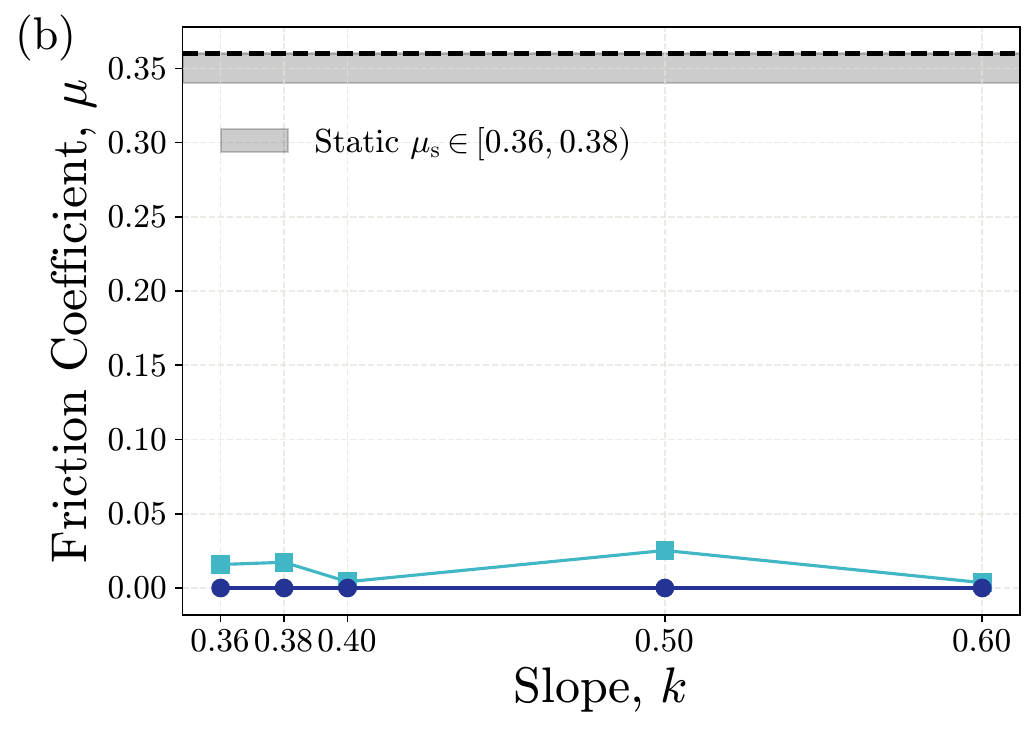}
    \hfill
    \includegraphics[width=0.49\linewidth]{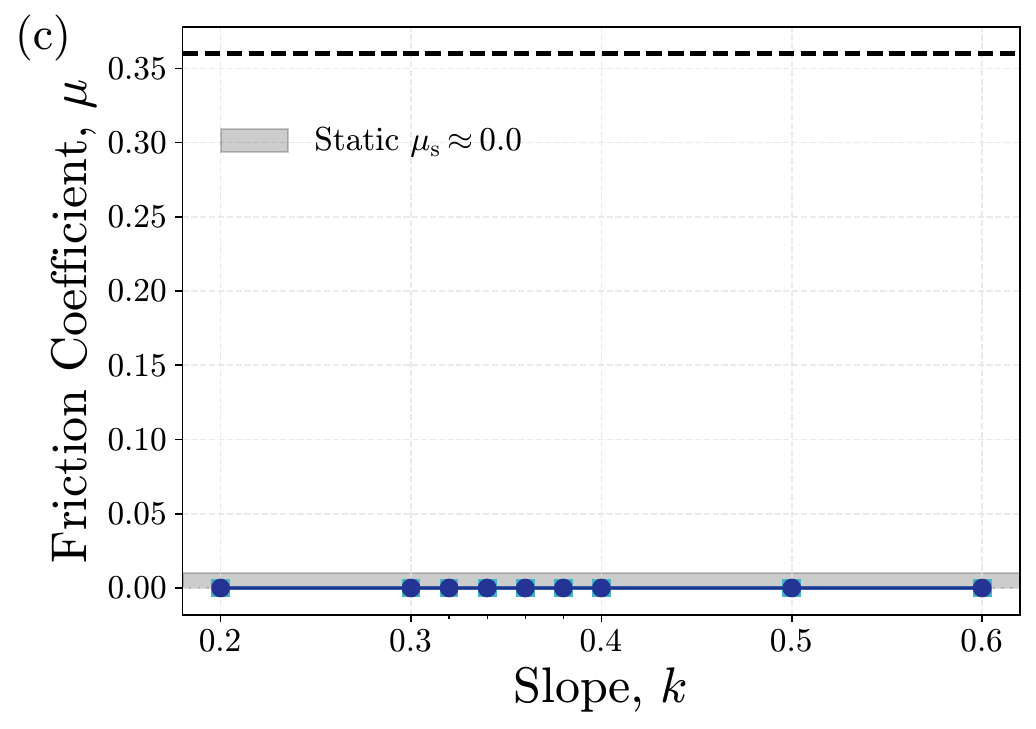}
    \hfill
    \includegraphics[width=0.49\linewidth]{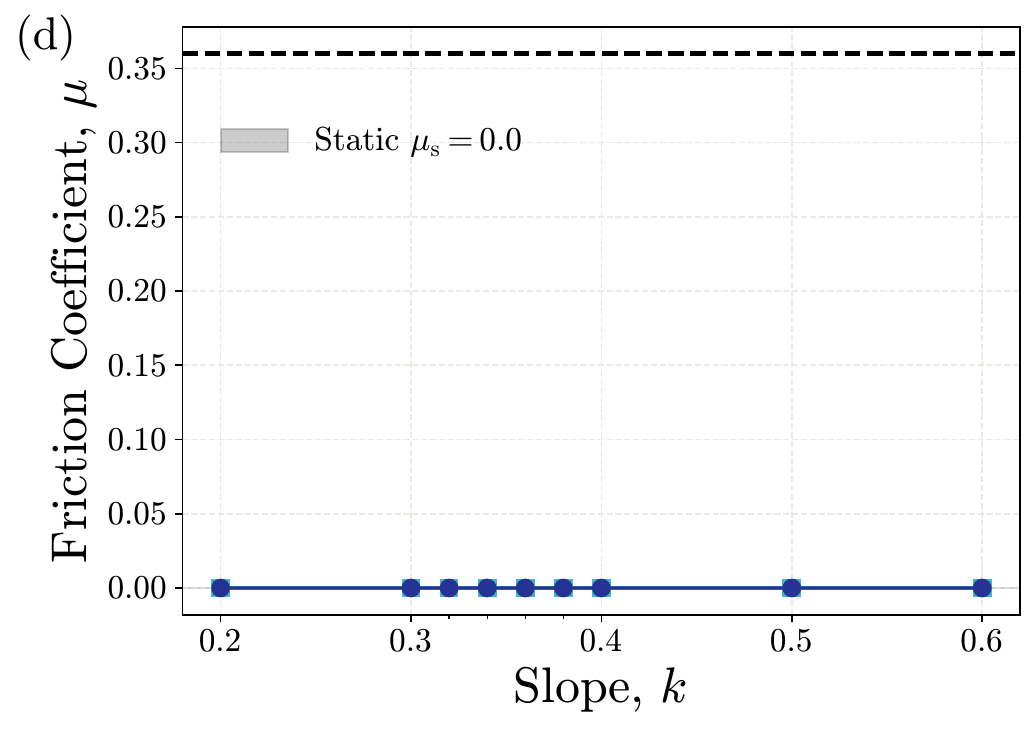}
    \caption{Effective friction coefficients for four combinations of surface roughness and normal dissipation. The kinematics-based $\mu_{k,\mathrm{eff}}$ and energy-based $\mu_{E,\mathrm{eff}}$ coefficients are shown for (a) rough surfaces with dissipative normal contacts ($e_n=0$); (b) rough surfaces with perfectly elastic collisions ($e_n=1$); (c) smooth surfaces with dissipative normal contacts; and (d) smooth surfaces with perfectly elastic collisions. Shaded regions indicate the the ranges of the effective static friction $\mu_\mathrm{s}$ inferred from the stick-to-slide transition, while the horizontal dashed line denotes the Coulomb friction reference $\mu_\text{p}=0.36$.}
    \label{fig:effective_friction_coefficient}
\end{figure}

To further identify the mechanisms responsible for the emergent frictional response, we examine the effects of surface roughness and normal contact dissipation separately and introduce two effective friction coefficients. First, a kinematics-based coefficient is obtained from the resisting force inferred from the tangential motion of the block. Following the formulation of Eq.~\eqref{analytical_acceleration}, we define
\begin{equation}\label{kinematics_based_mu}
    \mu_{k,\mathrm{eff}} = \frac{\int_0^t m(a_t+g\sin\theta)|v_t(\tau)|\,\rd\tau}{\int_0^t N(\tau)|v_t(\tau)|\,\rd\tau},
\end{equation}
where $N$ is the resultant normal contact force. Second, an energy-based coefficient is defined as
\begin{equation}\label{energy_based_mu}
    \mu_{E,\mathrm{eff}} = \frac{E_\mathrm{diss}}{\int_0^t N(\tau)|v_t(\tau)|\,\rd\tau},
\end{equation}
here $E_\mathrm{diss}$ is given by Eq.~\eqref{normal_energy_dissipation}. The former measures the total apparent resistance inferred from the block kinematics, whereas the latter quantifies the portion directly associated with irreversible normal contact dissipation.

Four cases are considered here: (A) rough surfaces with dissipative normal contact ($e_n=0$); (B) rough surfaces with perfectly elastic collision ($e_n=1$); (C) smooth surfaces with dissipative normal contact; and (D) smooth surfaces with perfectly elastic collision. As shown in Fig.~\ref{fig:effective_friction_coefficient}, Case A (Fig.~\ref{fig:effective_friction_coefficient}(a)) produces the strongest friction-like response, with both effective coefficients remaining finite over the sliding regime. When normal dissipation is removed while retaining roughness (Case B, Fig.~\ref{fig:effective_friction_coefficient}(b)), $\mu_{E,\mathrm{eff}}$ becomes negligible, as expected for a non-dissipative contact, whereas a small kinematics-based resistance and a finite stick-to-slide threshold remain. This indicates that geometric interlocking and momentum redirection alone can impede tangential motion, but are insufficient to reproduce the sustained kinetic resistance of the fully dissipative rough system. In contrast, retaining normal dissipation while removing roughness (Case C, Fig.~\ref{fig:effective_friction_coefficient}(c)) yields essentially zero effective friction. On a smooth surface, steady tangential motion produces little relative motion in the normal direction, and the normal damping mechanism is therefore not continuously activated. Case D (Fig.~\ref{fig:effective_friction_coefficient}(d)), with neither roughness nor dissipation, similarly approaches the frictionless limit.

These comparisons reveal complementary roles of the two mechanisms. Surface roughness converts macroscopic tangential motion into local normal motion and creates substantially more contact events through which dissipation can irreversibly remove (more) mechanical energy. Importantly, none of the Rough-DEM models considered above includes any local tangential contact force or any form of tangential dissipation. Thus, sustained kinetic-friction-like behavior emerges primarily from the coupling between roughness-induced normal motion and normal contact dissipation, rather than from either mechanism acting independently. Finally, our preliminary mechanisms-based analysis presented here suggests that a simple model may suffice to account for the complex connection between macroscopic friction and the microscopic interaction of two rough surfaces---a possibility we leave for a more in-depth study in future work.

\subsection{Recovery of Collective Frictional Behavior: Column Collapse Test}
\label{subsec:recover_collective_friction}

Having examined the emergence of (local) friction-like behavior at the two-body level, we next investigate whether geometric roughness can generate comparable effects in a collective granular system. The classical column collapse test is employed as a benchmark problem, where the final deposit morphology and angle of repose (AOR) provide direct measures of the bulk resistance to granular flow. By comparing the Rough-DEM results with those obtained from classical DEM simulations over a range of intergrain and grain-wall friction coefficients, we assess whether geometric surface roughness alone can generate collective frictional effects comparable to, or beyond, those produced by an explicit Coulomb friction law.

To keep our model's complexity comparable to that of classical DEM frameworks, here, the roughness profiles of both the grains and the confining walls are first restricted to a single sinusoidal mode, characterized by the roughness amplitude $h_\ast$ and wavelength $\lambda$, as defined in Eq.~\eqref{sinusoidal_roughness_profile}. This way, we replace the conventional friction parameters $(k_t,\mu_\text{p})$ with geometric parameters $(h,\lambda)$ without introducing additional degrees of freedom in the Rough-DEM. For each grain, the wavelength is controlled by the mode number $m$, which specifies the number of sinusoidal cycles along its smooth reference boundary. Accordingly, the nominal wavelength is defined as
\begin{equation}\label{wavelength_mode_relation}
    \lambda_i=\frac{L_i^{(0)}}{m},
\end{equation}
where $L_i^{(0)}$ denotes the perimeter of the unroughened grain. For the confining walls, the wavelength is determined using the median reference perimeter of all grains.

Simulations are performed in a two-dimensional domain of $300\times150\,\mathrm{mm}$, initially containing $1000$ grains confined behind a vertical gate at $x_0=45\,\mathrm{mm}$. The grain radii $r_i\in[0.8,1.2]\,\mathrm{mm}$. Before gate removal, the grains undergo gravitational settling and initial rearrangement to establish a mechanically stable packing. Upon gate removal, the global damping term characterized by $\eta$\footnote{Unlike the sliding block test in Section.~\ref{subsec:recover_local_friction}, nonzero global damping is introduced here to reduce the overall simulation time, without significantly contaminating the roughness-dependent trends in the final AOR.} presented in Eq.~\eqref{velocity_update} is activated to dissipate kinetic energy during column collapse. The principal simulation parameters, held constant unless otherwise specified, are summarized in Table~\ref{tab:basic_column_collapse_parameters}.

\begin{table}[!htbp]
    \setlength{\abovecaptionskip}{0cm}
    \setlength{\belowcaptionskip}{0cm}
    \centering
    \footnotesize
    \setlength{\tabcolsep}{4pt}
    \renewcommand{\arraystretch}{0.9}
    \caption{Principal numerical parameters for the column collapse simulations.}
    \label{tab:basic_column_collapse_parameters}
    \begin{tabular}{lccc}
        \toprule
        Parameter & Symbol & Value & Unit \\
        \midrule
        Initial grain arrangement & -- & $20\times50$ & -- \\
        Initial chamber width             & $x_0$     & $45$ & $\mathrm{mm}$ \\
        Grain radius  & $r_i$      & $[0.8,1.2]$ & $\mathrm{mm}$ \\
        Grain density & $\rho$ & $2.4\times 10^{-6}$ & $\mathrm{kg/mm^3}$ \\
        Normal contact stiffness & $k_n$ & 23536.08 & $N/m$ \\
        Coefficient of restitution & $c_n$      & $1.0$ & -- \\
        Global damping & $\eta$ & $50$ & $\mathrm{s^{-1}}$ \\
        Gravitational acceleration& $g$        & $-9806.7$ & $\mathrm{mm/s^2}$ \\
        Time step & $\Delta t$ & $1\times 10^{-5}$ & $\mathrm{s}$ \\
        Total running frames & $\ell$ & $150000$ & --\\
        \bottomrule
    \end{tabular}
\end{table}

\begin{figure}[!htbp]
    \setlength{\abovecaptionskip}{0cm}
    \setlength{\belowcaptionskip}{0cm}
    \centering
    \includegraphics[width=0.55\linewidth]{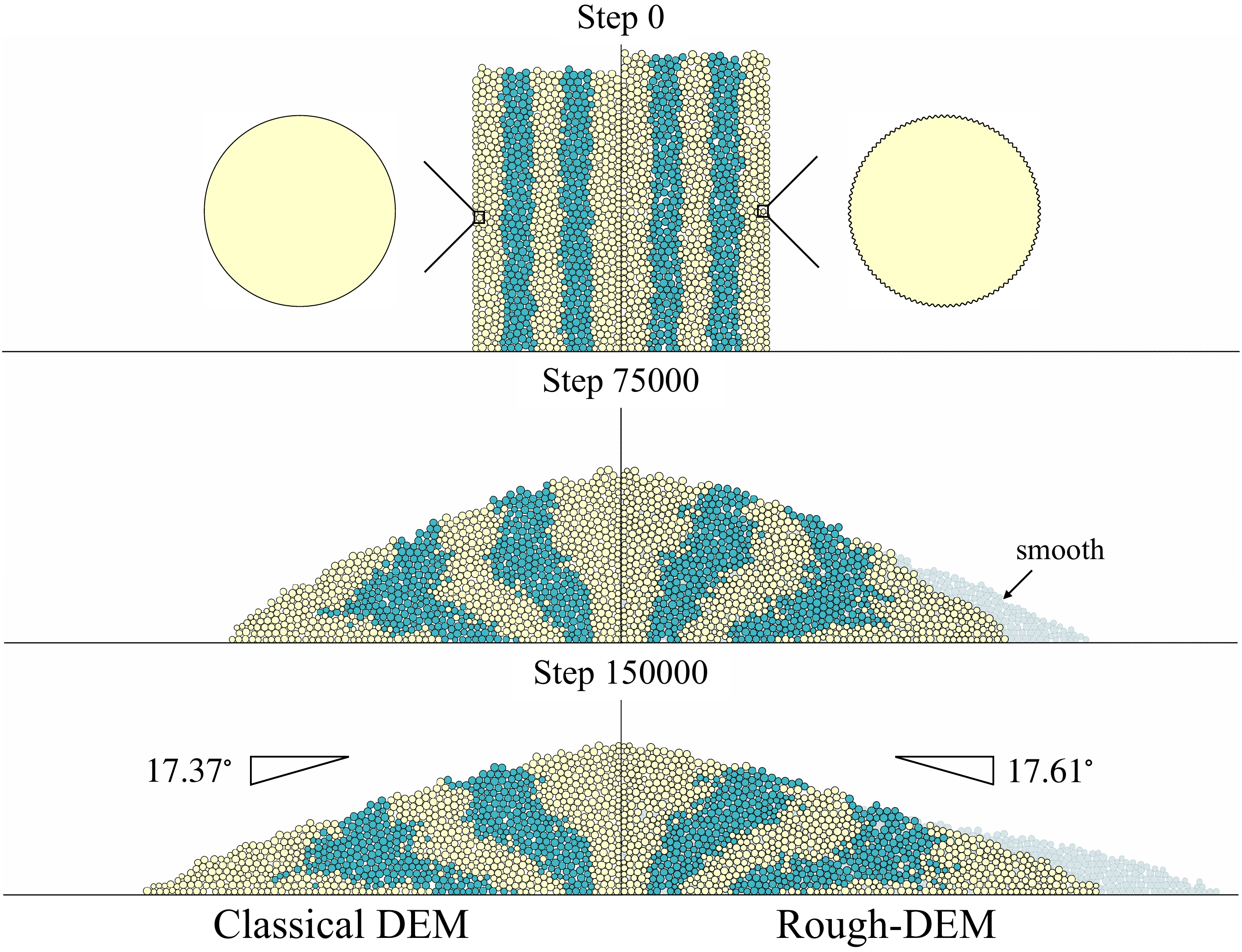}
    \caption{Comparison of granular column collapse of circular-shaped grains simulated using classical DEM (left, with $\mu_\text{p} =0.3$) and Rough-DEM with sinusoidal surface roughness (right) at different time steps. The magnified insets illustrate the grain geometries and surface characteristics of the two models. The lightly shaded particles overlaid on the right represent a reference Rough-DEM simulation with zero roughness amplitude (smooth grains), highlighting the additional resistance to granular spreading induced by surface roughness under identical global damping. The final angles of repose are indicated.}
    \label{fig:collapse_sinusoidal_recover}
\end{figure}

As a first example, we consider a roughness amplitude of $h_\ast=0.01\,\mathrm{mm}$ and a mode number of $m=100$. Fig.~\ref{fig:collapse_sinusoidal_recover} compares the final deposits obtained from the classical DEM and Rough-DEM, with magnified insets highlighting the differences in grain surface characteristics between the two models. The AOR is calculated from the slope of a robust Theil–Sen linear fit to the upper envelope of the final deposit, using points between $20\%$ and $80\%$ of the maximum deposit height to exclude the crest and toe regions. Despite the absence of an explicit Coulomb friction law, Rough-DEM produces a repose angle of $17.61^\circ$, closely matching the $17.37^\circ$ obtained from classical DEM with a friction coefficient of $\mu_\text{p}=0.3$. The lightly shaded particles overlaid on the right in Fig.~\ref{fig:collapse_sinusoidal_recover} represent a reference Rough-DEM simulation with zero roughness amplitude ($h=0$). At step 150000, the rough-grain assembly has already reached a stable configuration, whereas the zero-roughness reference remains in motion, with its kinetic energy not yet decayed to zero. The substantially greater spreading observed in the reference case, despite identical global damping, further highlights the role of surface roughness in resisting granular flow.

Next, to demonstrate the versatility of Rough-DEM in modeling arbitrary surface-roughness geometries, we consider circular grains with roughness described by the random Fourier series defined in Eq.~\eqref{fourier_series_roughness}. Each grain surface is constructed using 180 Fourier modes with randomly assigned phases, while the roughness amplitude parameter $h_\ast$ and spectral decay exponent $p$ are sampled from $[0.005,0.015]$ mm and $[0.5,0.8]$, respectively. The confining walls are similarly assigned random Fourier roughness, with 8000 modes, an amplitude parameter of $0.02$ mm, and a decay exponent of $0.5$. The remaining numerical settings follow those presented in Table~\ref{tab:basic_column_collapse_parameters}.

\begin{figure}[!htbp]
    \setlength{\abovecaptionskip}{0cm}
    \setlength{\belowcaptionskip}{0cm}
    \centering
    \includegraphics[width=0.5\linewidth]{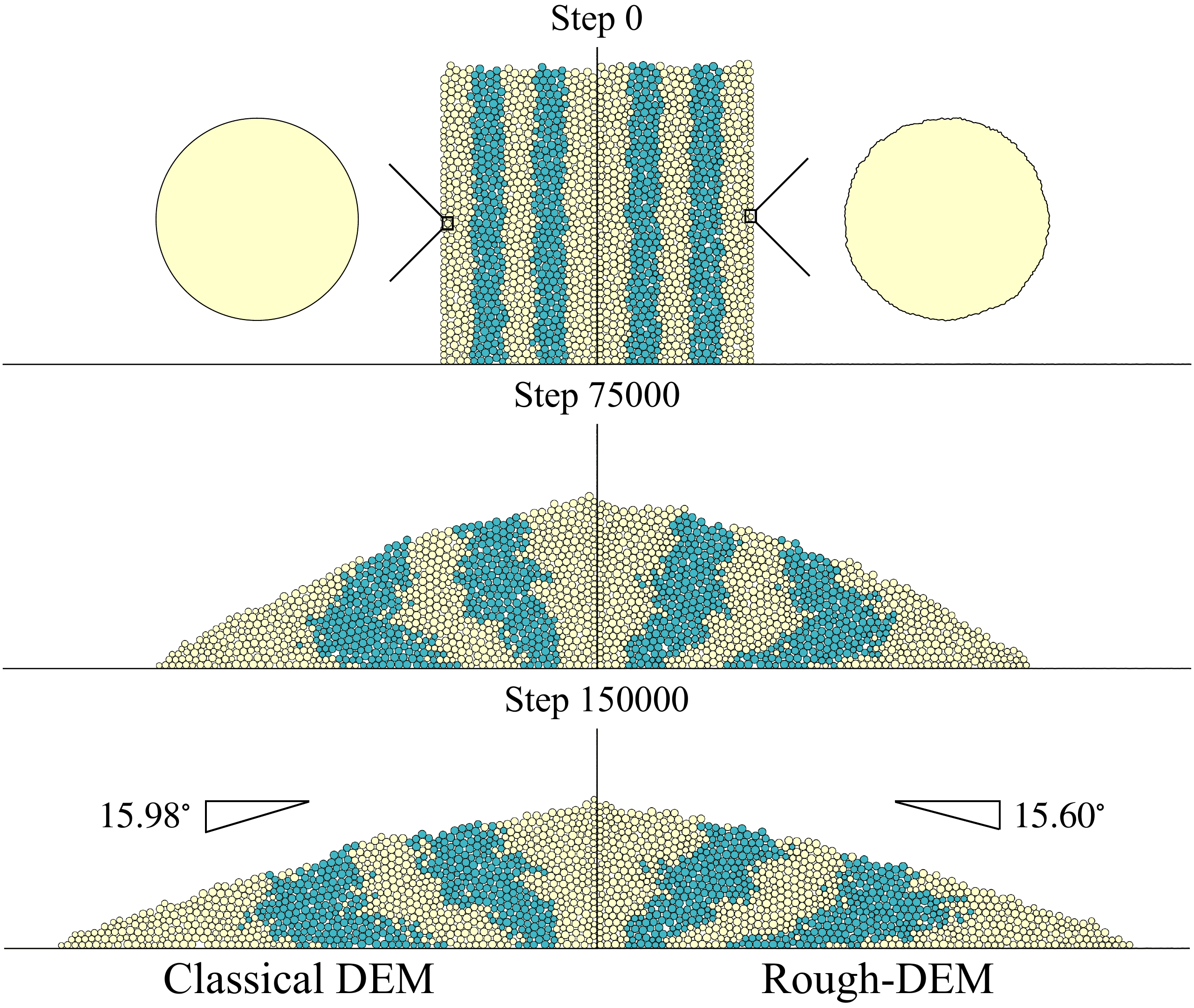}
    \caption{Comparison of granular column collapse of circular-shaped grains simulated using classical DEM (left, with $\mu_\text{p} = 0.12$) and Rough-DEM with Fourier-series surface roughness (right) at selected time steps. The magnified insets highlight the differences in grain surface geometry between the two models. The final angles of repose are indicated.}
    \label{fig:collapse_circular_recover}
\end{figure}

Fig.~\ref{fig:collapse_circular_recover} compares the column collapse results obtained from Rough-DEM with random Fourier roughness and classical DEM with a friction coefficient of $\mu_\text{p}=0.12$. The resulting AOR are $15.60^\circ$ and $15.98^\circ$, respectively, showing close agreement in the final deposit morphology. This result further demonstrates that the collective friction-like behavior generated by geometrically resolved normal contacts is not restricted to simple sinusoidal roughness but can also emerge from more complex, multiscale surface geometries.

\subsection{Beyond the Saturation of Coulomb Friction}
\label{subsec:beyond_classical_dem}

In this subsection, we further investigate whether geometric roughness can generate macroscopic resistance beyond what is attainable by increasing Coulomb friction in classical DEM models. Both models still adopt the baseline simulation setup summarized in Table~\ref{tab:basic_column_collapse_parameters}. For classical DEM, a common Coulomb friction coefficient $\mu_\text{p}$ is prescribed for particle–particle and particle–wall contacts and systematically varied from $0.2$ to $5.0$, while the remaining parameters are held constant. For Rough-DEM, we adjust the sinusoidal roughness amplitude $h_\ast$ and mode number $m$, which is related to the nominal wavelength through Eq.~\eqref{wavelength_mode_relation}, to explore whether geometric roughness can produce a larger deposit angle than classical DEM. The resulting roughness parameters are $h_\ast=0.032$ and $m=200$, with the amplitude corresponding to approximately $2.7-4\%$ of the particle radii ($r_i\in[0.8,1.2]$)\footnote{It is in our opinion that $2.7-4\%$ is small enough to be considered as surface roughness; it is unclear however and a quite interesting question to us in terms of exactly how small the surface topography (with respect to the typical grain size) should be to be considered as surface roughness, rather than as being part of the base shape of the grain. This question is conceptually similar to the limit of separation of length scales commonly assumed in classical, continuum-based multiscale analysis of solids.}

\begin{figure}[!htbp]
    \setlength{\abovecaptionskip}{0cm}
    \setlength{\belowcaptionskip}{0cm}
    \centering
    \includegraphics[width=0.55\linewidth]{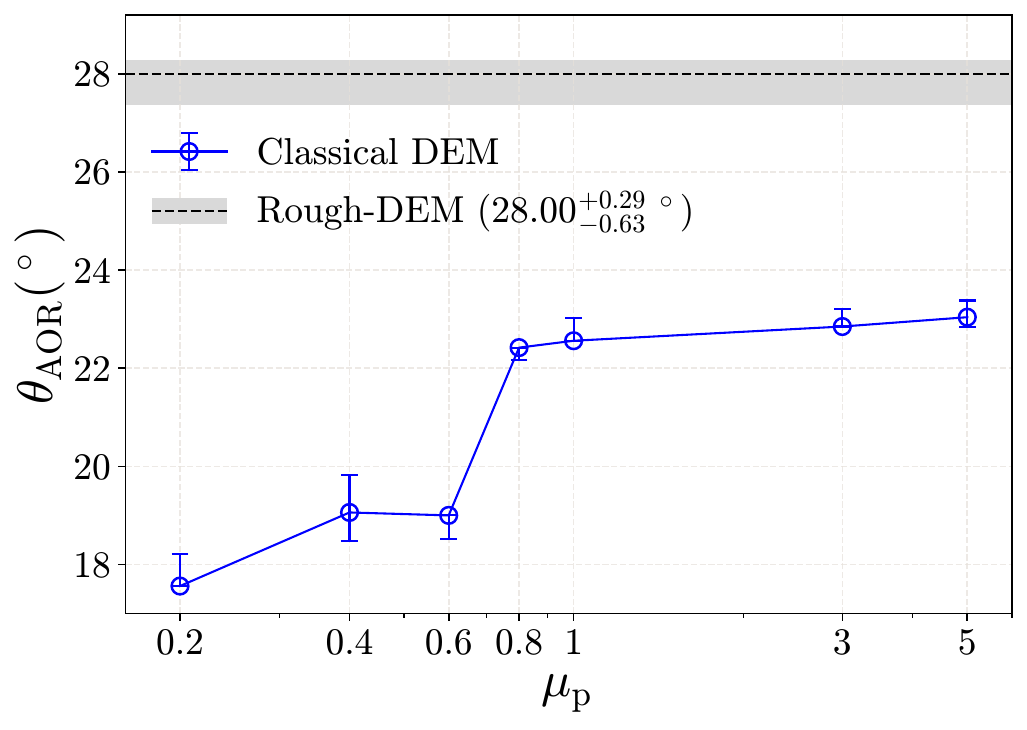}
    \caption{Final deposit angle $\theta_{\mathrm{AOR}}$ as a function of the Coulomb friction coefficient $\mu_{\mathrm{p}}$ in column-collapse simulations using classical DEM. Markers and the horizontal solid line represent classical DEM and Rough-DEM results, respectively, fitted over $20-80\%$ of the final pile height. Vertical whiskers and the shaded band indicate the corresponding sensitivity to alternative fitting intervals ($10-90\%$ and $30-70\%$).}
    \label{fig:angle_of_repose_vs_mu}
\end{figure}

As shown in Fig.~\ref{fig:angle_of_repose_vs_mu}, the deposit angle predicted by classical DEM generally increases with $\mu_\text{p}$ and appears to saturate at higher friction coefficients, despite local non-monotonic variations. Notably, this trend persists even when $\mu_\text{p}$ is increased to 5.0, whereas sliding friction coefficients above 0.5 are already considered unrealistically large in common modeling scenarios for granular materials \cite{jenkins2026unphysical}.

To assess the sensitivity to the measurement procedure, the deposit angle is evaluated using three fitting intervals spanning $10-90\%$, $20-80\%$, and $30-70\%$ of the final pile height. The maximum observed classical DEM angle among the cases shown is $23.38^\circ$, whereas the Rough-DEM achieves $28.00^\circ$ using the $20-80\%$ interval, with values ranging from $27.37^\circ$ to $28.29^\circ$ across the three fitting intervals. Even the lowest Rough-DEM value exceeds the maximum classical DEM value in this comparison, demonstrating the potential of explicitly resolved surface roughness to capture strong geometric interlocking and enhanced macroscopic frictional resistance.

The apparent saturation in classical DEM suggests that further increasing the Coulomb friction coefficient becomes progressively less effective in modifying the macroscopic deposit geometry. While increasing $\mu_\text{p}$ raises the allowable tangential contact force, it does not alter the local contact geometry. In contrast, the Rough-DEM explicitly resolves surface asperities, allowing geometric interlocking and resistance to relative tangential motion and particle rotation through normal contact interactions. This distinct mechanism highlights the potential of the Rough-DEM to capture the behavior of granular materials exhibiting strong interlocking and pronounced macroscopic frictional resistance.

\begin{figure}[!htbp]
    \setlength{\abovecaptionskip}{0cm}
    \setlength{\belowcaptionskip}{0cm}
    \centering
    \includegraphics[width=0.85\linewidth]{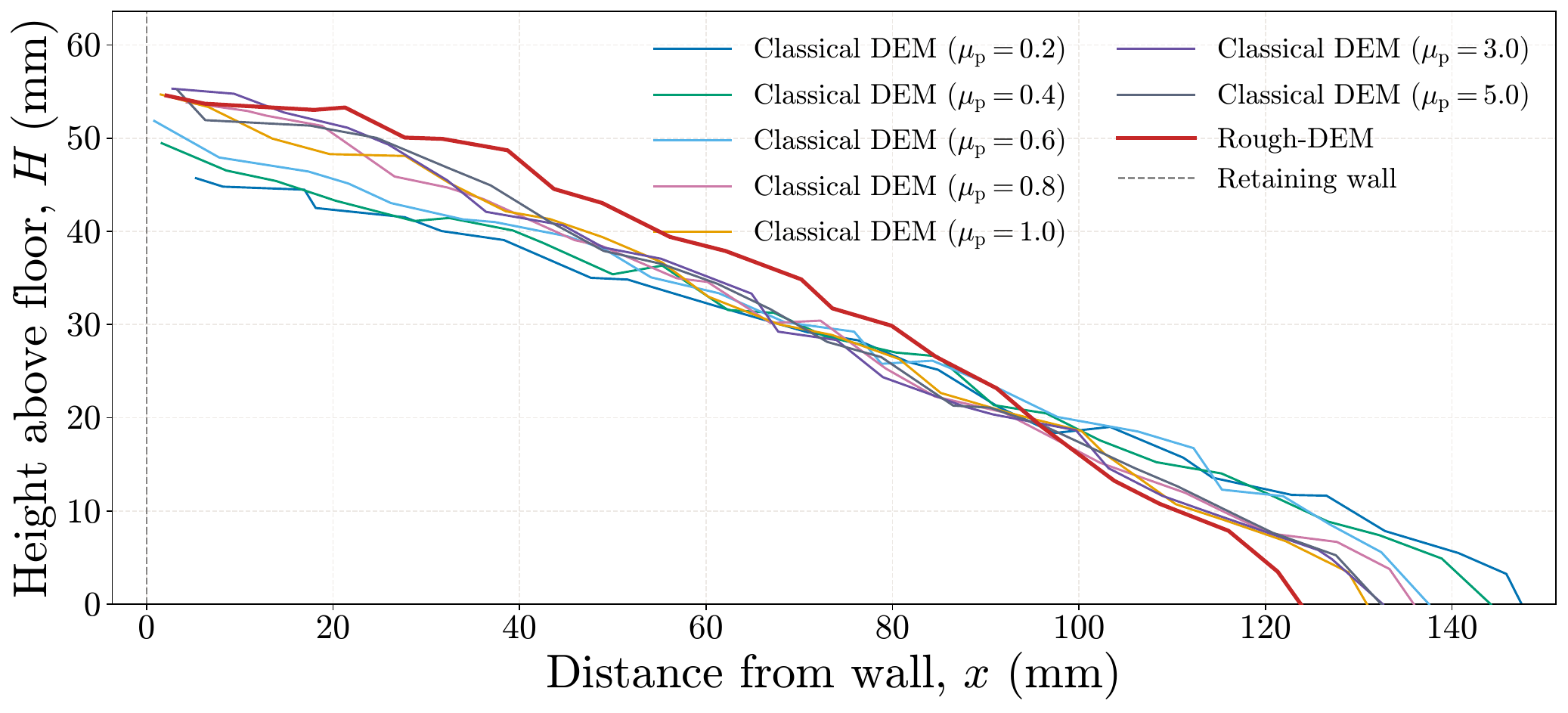}
    \caption{Final deposit profiles obtained from column-collapse simulations using classical DEM with different Coulomb friction coefficients and the Rough-DEM.}
    \label{fig:final_deposit_profile}
\end{figure}

The final deposit profiles in Fig.~\ref{fig:final_deposit_profile} further illustrate this difference in macroscopic behavior. Compared with the classical DEM, the Rough-DEM produces a more compact deposit with a higher free surface over much of the upstream region and a shorter runout distance, consistent with its larger deposit angle.

\section{Numerical Implementation and Acceleration Strategies}
\label{sec:acceleration_strategy}
The computational cost of grain-level algorithms is often prohibitive; further, it scales with both the number of grains and the number of local contacts and becomes even more substantial when small-scale roughness is explicitly resolved. To improve the computational efficiency while retaining the ability to investigate rough-surface interactions, this study adopts several acceleration strategies, including local window refinement (i.e., the FE$^2$-like algorithm, see also Fig.~\ref{fig:hierarchical_contact}(b) and (c)), GPU acceleration, and a neural network surrogate for signed distance evaluation. Since Section.~\ref {subsec:contact_detection} already discussed the local-window approach for contact detection that accounts for detailed surface roughness, this section focuses on implementing the other two acceleration strategies. The previously introduced column collapse test is also used to evaluate both the computational acceleration achieved by these strategies and the predictive accuracy of the proposed SDF surrogate model.

\subsection{GPU-Accelerated Batched Contact Evaluation}
\label{subsec:gpu_implementation}
A major computational cost of the present method arises from the narrow-phase evaluation of roughness contacts. For each candidate pair, the solver must reconstruct roughness geometry, evaluate signed distances, and integrate the resulting nodal forces and moments. Performing all these operations sequentially for individual grain pairs and surface nodes introduces substantial Python overhead and does not fully utilize the parallelism available on modern GPU. The roughness-scale contact calculation is therefore reformulated as a batched tensor pipeline using PyTorch, with per-pair and per-node loops replaced by tensor dimensions.

The implementation employs a GPU-resident computational pipeline, with CPU operations primarily used for initialization and data recording. During initialization, static geometric data, including the homogeneous control points, Fourier coefficients, body-frame boundary nodes, reference perimeters, bounding radii, grain areas, and moments of inertia, are prepared on CPU and transferred to GPU memory for subsequent computations. At each time step, nearest-neighbor search and bounding-sphere filtering identify candidate grain pairs for batched rough-contact evaluation, followed by GPU-based time integration using the resulting forces and moments. To support regular tensor batching, all grains share the same NURBS degree, knot vector, weights, and boundary discretization, while their control-point coordinates and roughness coefficients may differ.

To describe the specific tensor organization, let $N_g$ denote the total number of grains, $N_c$ the number of candidate pairs processed in each GPU chunk, $N_b$ the number of base boundary nodes per grain, and $N_r$ the number of locally refined nodes per contact window. The chunk size $N_c$ is limited to control peak GPU memory consumption. Grain centroids and translational velocities are represented by tensors in $\mR^{N_g\times2}$, while their orientations and angular velocities are stored in $\mR^{N_g}$. After rigid-body transformation (Eq.~\eqref{rigid_body_transformation}), the base nodes of all grains form a tensor in $\mR^{N_g\times N_b\times 2}$. For each candidate pair, the NURBS parameters sampled within the master and slave contact windows are organized into tensors of size $N_c\times N_r$. These parameters are flattened for vectorized NURBS evaluation, yielding the coordinate and normal tensors
\begin{equation}
\vX_\mathrm{m},\vX_\mathrm{s},\vN_\mathrm{m},\vN_\mathrm{s}\in\mR^{N_c\times N_r\times 2},
\end{equation}
where $\vX_\mathrm{m}$ and $\vX_\mathrm{s}$ contain the coordinates sampled from the master and slave base curves $\vC(u)$, respectively, while $\vN_\mathrm{m}$ and $\vN_\mathrm{s}$ contain their corresponding unit normals. The Fourier perturbations evaluated at the same local parameter coordinates are stored in tensors $\vh_\mathrm{m},\vh_\mathrm{s}\in\mR^{N_c\times N_r}$. Finally, the locally reconstructed rough boundaries are obtained according to Eq.~\eqref{roughness_description}, which can be expressed in tensor form as
\begin{equation}
    \vX_\mathrm{m}^\mathrm{rough} = \vX_\mathrm{m}+\left(\vh_\mathrm{m}\otimes\mathbf{1}_2\right)\odot\vN_\mathrm{m},\quad\vX_\mathrm{s}^\mathrm{rough} = \vX_\mathrm{s}+\left(\vh_\mathrm{s}\otimes\mathbf{1}_2\right)\odot\vN_\mathrm{s},
\end{equation}
where $\mathbf{1}_2=[1,1]^\rT\in\mR^2$, $\otimes$ denotes the tensor product that expands the roughness perturbation tensor along the spatial dimension, and $\odot$ denotes element-wise multiplication.

The locally reconstructed slave boundaries are augmented with closing arcs to form closed polygons with a uniform number of vertices, represented as tensors in $\mR^{N_c\times N_s\times 2}$, where $N_s=N_r+N_a+1$ and $N_a$ is the number of nodes used to discretize each closing arc. The batched signed distance calculation therefore takes the master rough nodes, $\vQ\in\mR^{N_c\times N_r\times2}$, and the corresponding slave polygons, $\vS\in\mR^{N_c\times N_s\times2}$, as its major inputs. It returns signed distances in $\mR^{N_c\times N_r}$ and outward normals in $\mR^{N_c\times N_r\times2}$.

To further accelerate the signed-distance calculations, two fused GPU kernels are implemented using Triton within the PyTorch pipeline. The first kernel performs single-precision (FP32) screening to identify potential contact nodes and nearest-node candidates using a geometry-scaled tolerance. The second kernel recomputes the relevant signed distances and, when required, outward normals in double precision (FP64). Both kernels process query–edge interactions in tiles, avoiding the construction of large intermediate distance tensors. The Triton kernels are validated against the PyTorch implementation at initialization, with an automatic fallback if the validation fails.

The resulting signed distances and contact normals are then used to calculate nodal contact forces, which are summed to obtain pair-level resultant forces in $\mathbb{R}^{N_c\times2}$. Indexed accumulation is then used to assemble the total grain forces and moments in $\mR^{N_g\times2}$ and $\mR^{N_g}$, respectively.

Grain–wall contacts are handled through an analogous GPU-batched procedure for grains passing the wall-proximity test, using the same mixed-precision signed-distance evaluation strategy. The assembled grain forces and moments are used for vectorized time integration, after which the updated states and accelerations are transferred to the CPU in a single packed operation for trajectory recording.

\subsection{Neural Network Surrogate for Signed Distance Function}
\label{subsec:nn_surrogate}
Although the batched GPU implementation substantially reduces the overhead associated with sequential contact evaluation, it does not alter the underlying geometric workload. For each candidate pair, the exact formulation still requires the local reconstruction of both rough surfaces, the construction of a closed slave polygon, and the evaluation of $N_r$ master query points against $N_s$ slave boundary segments. The computational and memory costs of roughness-scale signed distance calculation therefore scale approximately as $\mO\left(N_cN_rN_s\right)$. To reduce this remaining geometry-dependent cost, a neural surrogate is introduced to replace the exact roughness-scale signed distance evaluation while retaining the same broad-phase detection, master-boundary refinement, contact law, and force-and-moment assembly.

In general, computing the signed distances from a set of query points $\vq$ to a given geometry $\mathcal{G}$, represented by a curve or surface, requires evaluating an appropriate signed distance procedure, denoted by $\mathrm{SDF}(\cdot)$. Common approaches include interpolation from a precomputed level-set field and closest-point projection onto the geometry. Specifically,
\begin{equation}\label{signed_distance_general}
    \vd = \mathrm{SDF}\left(\vq\,|\,\mathcal{G}\right).
\end{equation}
This evaluation can become computationally expensive when both the number of query points and the number of nodes used to discretize the known geometry are large, particularly if multiple iterations are required to obtain a converged signed distance solution for each query point. In the context of surface roughness, the challenge is further amplified by the significant distinction between roughness and overall base shape. To address this issue, we approximate the roughness-scale signed distance function in Eq.~\eqref{signed_distance_general} using a pretrained neural network $\mathcal{N}_\theta$ that encodes the information of base geometry and local roughness, i.e.,
\begin{equation}\label{signed_distance_nn}
    \hat{\vd} = \mathcal{N}_\theta\left(\vx_\mathrm{local}\mid\vz_g\right),
\end{equation}
where $\hat{\vd}$ denote the scaled signed distance prediction. Here, $\vx_\mathrm{local}$ denotes the local input vector associated with each query point, including its coordinates and relative position features, whereas $\vz_g$ is a latent descriptor of the grain geometry. The two network inputs serve distinct roles in Eq.~\eqref{signed_distance_nn}: $\vx_\mathrm{local}$ specifies where the signed distances are evaluated, whereas $\vz_g$ identifies the grain geometry and therefore determines which signed distance field to approximate.

The local input vector is constructed from the position of each query point relative to the unperturbed boundary of the slave grain. Let $\vq$ denote the query coordinates expressed in the body-fixed frame of the slave grain. The signed distance $d_\mathrm{b}$ from $q$ to the base boundary and the arc-length coordinate $t_\mathrm{b}$ of the corresponding closest boundary point are first computed. The latter is then used to evaluate the local Fourier roughness height $h(t_\mathrm{b})$, together with its first- and second-derivative features, $h^\prime(t_{\mathrm{b}})$ and $h^{\prime\prime}(t_{\mathrm{b}})$, which characterize the local roughness slope and curvature. The local input vector is therefore defined as
\begin{equation}\label{NN_local_input_vector}
    \vx_\mathrm{local} = \left[\vq,d_\mathrm{b},t_\mathrm{b},h(t_\mathrm{b}),h^\prime(t_\mathrm{b}),h^{\prime\prime}(t_\mathrm{b})\right].
\end{equation}

Particular attention is given to the scaling of the distance-related quantities. The roughness-scale signed distance samples relevant to contact initiation and penetration are highly concentrated within a narrow band around zero. Direct regression in physical units would therefore place these mechanically important variations within a small numerical range relative to the other network features, leading to an ill-conditioned learning problem and making it more difficult for the network to resolve near-contact behavior accurately. To expand this range, a prescribed characteristic distance scale, $d_\mathrm{scale}$, is introduced. This treatment is conceptually consistent with the scaling strategy used in learned 3D shape representations such as DeepSDF \cite{park2019deepsdf}. The network is trained to predict the scaled signed distance $\hat{\vd}$, which is then converted back to the physical unit through
\begin{equation}
    \tilde{\vd} = d_\mathrm{scale}\hat{\vd},
\end{equation}
where $\tilde{\vd}$ is the network approximation of the signed distance $\vd$ defined in Eq.~\eqref{signed_distance_general}.

\begin{figure}[!htbp]
    \setlength{\abovecaptionskip}{0cm}
    \setlength{\belowcaptionskip}{0cm}
    \centering
    \includegraphics[width=0.85\linewidth]{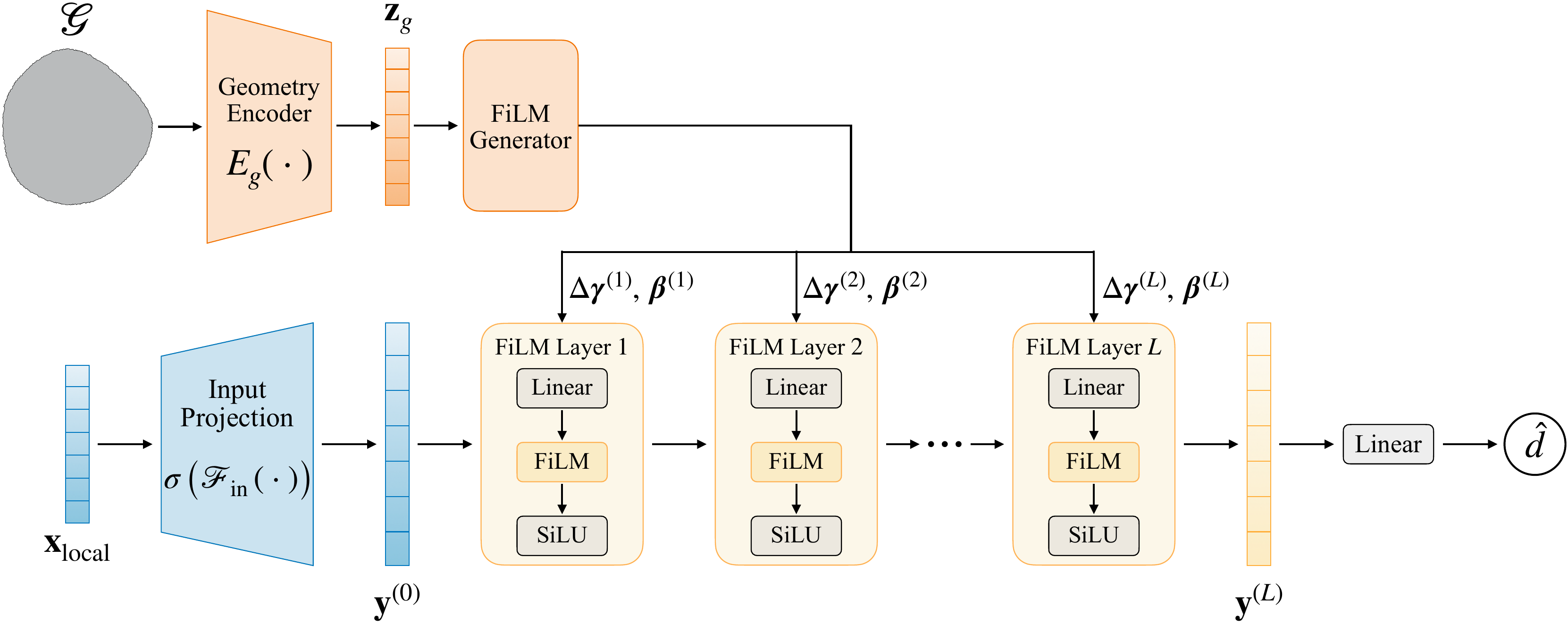}
    \caption{Architecture of the FiLM-SDF network surrogate. The local input features $\vx_\mathrm{local}$ are transformed into the initial hidden representation $\vy^{(0)}$ through a linear projection $\CF_\mathrm{in}$ followed by the SiLU activation $\sigma$. The grain geometry $\mathcal{G}$ is encoded into the latent geometry descriptor $\vz_g$, which is used by the FiLM generator to produce layer-wise modulation parameters $\{\Delta\vgamma^{(l)},\vbeta^{(l)}\}_{l=1}^L$. Each hidden layer consists of a linear transformation, FiLM modulation, and SiLU activation. The final linear layer maps $\vy^{(L)}$ to the predicted signed distance $\hat{\vd}$.}
    \label{fig:film_sdf_architecture}
\end{figure}

After constructing the scaled local input features and the geometry descriptor, a straightforward way to incorporate the grain geometric condition is to concatenate $\vx_\mathrm{local}$ and $\vz_g$ at the network input. Such input-level concatenation, however, provides the geometry descriptor with only an indirect pathway for influencing the subsequent hidden representations. To introduce the geometric condition more explicitly throughout the network and force it to output different signed distance predictions under different grain geometries, we employ the feature-wise linear modulation (FiLM) \cite{perez2018film}. A typical FiLM transformation is written as
\begin{equation}\label{film_layer_expression}
    FiLM(\vu^{(l)}\,|\,\Delta\vgamma^{(l)},\vbeta^{(l)}) = \left(1+\Delta\vgamma^{(l)}\right)\odot\vu^{(l)}+\vbeta^{(l)},
\end{equation}
where $\Delta\vgamma^{(l)}$ and $\vbeta^{(l)}$ are feature-wise scale increments and shifts corresponding to layer $l$, respectively. In this work, these two coefficients are generated from the grain-geometry descriptor $\vz_g$. For different grain geometries, $\vz_g$ directly modulates the hidden representations through $\Delta\vgamma^{(l)}$ and $\vbeta^{(l)}$, thereby enabling a single shared network to approximate a family of signed distance fields corresponding to different grain geometries and roughness realizations.

Specifically, this modulation process can be expressed as
\begin{subequations}\label{film_layer_process}
    \begin{align}
        &\vy^{(0)} = \sigma\left(\CF_\mathrm{in}\left(\vx_\mathrm{local}\right)\right);\\
        &\left[\Delta\vgamma^{(l)},\vbeta^{(l)}\right] = \CF_\mathrm{gen}^{(l)}\left(\vz_g\right),\quad l=1,\,\cdots,L;\\
        &\vy^{(l)} = \sigma\left[FiLM\left(\CF^{(l)}\left(\vy^{(l-1)}\right)\,|\,\Delta\vgamma^{(l)},\vbeta^{(l)}\right)\right],
    \end{align}
\end{subequations}
where $\sigma(\cdot)$ denotes the SiLU activation function, and $\CF_\mathrm{in},\CF_\mathrm{gen}^{(l)}$, and $\CF^{(l)}$ denote the linear transformations used to project the local input features, generate the layer-wise FiLM coefficients, and transform the hidden features before FiLM modulation in the $l$-th FiLM layer, respectively. The FiLM-modulated neural representation of the signed distance function is referred to as the FiLM-SDF network, whose architecture is illustrated in Fig.~\ref{fig:film_sdf_architecture}.

The FiLM-SDF network is trained using a weighted regression loss together with an auxiliary sign-consistency term. For the $i$-th sample, let $d_i$ denote the physical signed distance measured in the presence of roughness, $\bar{d}_i = d_i/d_\mathrm{scale}$ the corresponding scaled target, and $\hat{d}_i$ the scaled network prediction. The total loss is defined as
\begin{equation}\label{loss_function_nn}
    \CL = \frac{1}{N}\sum_{i=1}^Nw_i\rho_\beta\left(\bar{d}_i-\hat{d}_i\right)+\frac{\lambda_\mathrm{sign}}{N_\mathrm{nz}}\sum_{i:\bar{d}_i\neq0}\log\left(1+\exp\left(-\frac{\mathrm{sign}(\bar{d}_i)\hat{d}_i}{m}\right)\right),
\end{equation}
where $N$ is the number of samples in the batch and $N_\mathrm{nz}$ is the number of samples with nonzero target signed distance used for sign-consistency check. The function $\rho_\beta$ denotes the Smooth L1 loss,
\begin{equation}\label{smooth_l1_loss}
    \rho_\beta(e) = \begin{cases}
    \frac{e^2}{2\beta}, &|e|\leq\beta,\\
    |e|-\frac{\beta}{2}, &|e|>\beta,
    \end{cases}
\end{equation}
with $\beta$ controlling the transition between the quadratic and linear regimes of the regression loss.

To emphasize samples near the contact boundary, the regression term is weighted according to the true physical distance:
\begin{equation}\label{weight_loss}
    \tilde{w}_i = 1+\left(w_\mathrm{peak}-1\right)\exp\left[-\frac{1}{2}\left(\frac{|d_i|}{b}\right)^2\right],\quad w_i = \frac{\tilde{w}_i}{\frac{1}{N}\sum_{j=1}^N\tilde{w}_j}.
\end{equation}

Here, $w_\mathrm{peak}$ specifies the maximum weight assigned at the grain surface ($d_i=0$), while $b$ controls the width of the emphasized near-contact region. The batch-wise normalization enforces a unit mean weight and prevents substantial variations in the overall magnitude of the regression loss across batches. The sign-consistency term penalizes predictions with an incorrect sign, with $\lambda_\mathrm{sign}$ controlling the relative contribution of this auxiliary term and $m$ serving as a positive scaling parameter that determines the sharpness of the sign penalty in the scaled-distance space.

\subsection{Neural Performance for Predicting Collective Behavior}
\label{subsec:nn_performance}
The network parameters are selected following the FiLM-SDF architecture introduced in Fig.~\ref{fig:film_sdf_architecture}. The hidden feature dimension is set to $H=256$, corresponding to the dimension of the hidden representations $\mathbf{y}^{(l)}$ throughout the input projection and FiLM layers. The geometry encoder $E_g(\cdot)$ maps the grain geometry descriptors $\mathcal{G}$ to a latent vector $\vz_g\in\mR^{128}$, with an encoder multilayer perceptron (MLP) depth of four. A learnable embedding of dimension 64 is used to represent the discrete Fourier-mode count $N_f$, with the maximum number of Fourier modes set to 200. The FiLM-conditioned network contains $L=5$ layers, where the latent geometry descriptor $\vz_g$ is used by the FiLM generator to produce the layer-wise modulation parameters $\{\Delta\boldsymbol{\gamma}^{(l)},\boldsymbol{\beta}^{(l)}\}_{l=1}^{L}$. The network is trained using the loss function defined in Eqs.~\eqref{loss_function_nn}-\eqref{weight_loss}. The physical roughness-scale signed distance is scaled by $d_{\mathrm{scale}}=0.015$ before regression. For the Smooth L1 term, the transition parameter is set to $\beta=0.05$. To emphasize accuracy near the contact boundary, the Gaussian weighting function uses $w_{\mathrm{peak}}=5$ and a characteristic width $b=0.01$. The sign-consistency contribution is weighted by $\lambda_{\mathrm{sign}}=0.05$, with $m=0.05$ controlling the sharpness of the sign penalty. Finally, the training is performed using the AdamW optimizer with an initial learning rate of $2\times10^{-4}$ and a weight decay of $10^{-5}$. At convergence, the training and validation losses are 0.081684 and 0.081797, respectively.

With the trained FiLM-SDF surrogate, we are ready to incorporate it into the original contact computation process to replace the costly evaluation of roughness-scale signed distances. This choice reflects a trade-off between computational efficiency and accuracy. In principle, a neural surrogate could replace a larger portion of the contact calculation and potentially provide greater acceleration. However, increasing the extent of neural approximation also introduces additional prediction errors, which could accumulate during the entire simulation. Therefore, we restrict the surrogate to the roughness-scale SDF evaluation, which represents one of the most computationally expensive geometric operations while retaining the physics-based contact formulation.

The computational efficiency of the FiLM-SDF surrogate is further improved through a grain-specific lookup-table strategy. Since each grain's geometry and surface roughness remain fixed in its body frame, a two-dimensional grid is constructed in terms of the normalized base-boundary arc-length coordinate $t_b$ and signed distance $d_b$. At each grid point, the trained network is evaluated using the complete set of local input features presented in Eq.~\eqref{NN_local_input_vector}, including the Fourier roughness height and its derivatives. The predicted signed distances and their partial derivatives are precomputed within a narrow band around the grain boundary and stored for subsequent use. During contact evaluation, a fused Triton kernel performs base-boundary projection and bicubic interpolation to approximate the signed distances, while the corresponding contact normals are reconstructed from the interpolated derivatives using the chain rule. The lookup tables are validated against direct network predictions and refined when necessary, with unsuitable queries handled by the original network or excluded when contact can be ruled out. This strategy reduces repeated NN inference without modifying the underlying contact formulation.

\begin{figure}[!htbp]
    \setlength{\abovecaptionskip}{0cm}
    \setlength{\belowcaptionskip}{0cm}
    \centering
    \includegraphics[width=0.5\linewidth]{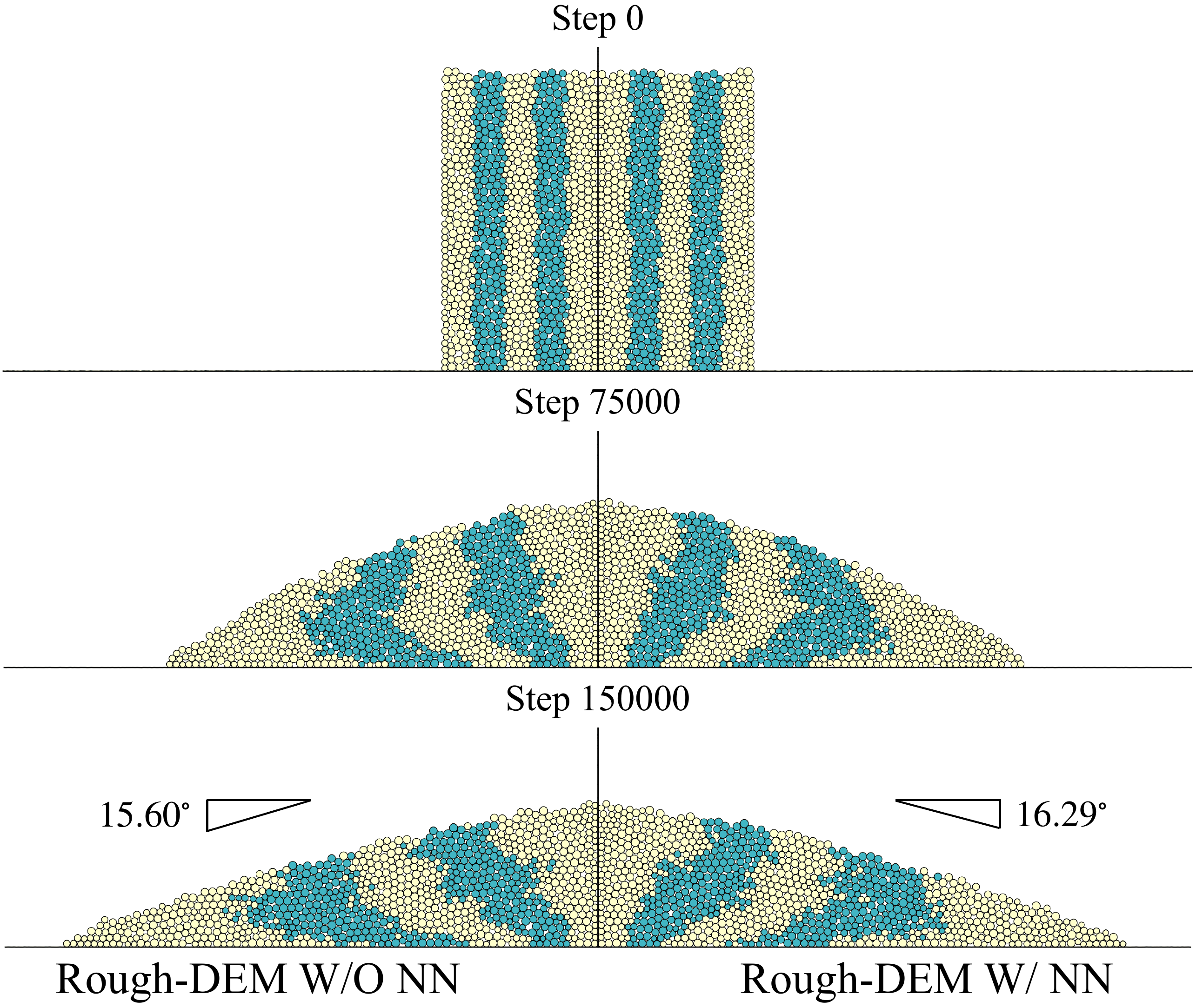}
    \caption{Comparison of column collapse simulations of circular-shaped grains using the proposed Rough-DEM model without (left) and with (right) the NN surrogate. The final repose angles are indicated.}
    \label{fig:collapse_circular_NN}
\end{figure}

To assess whether the computational acceleration is achieved without significantly altering the resulting granular dynamics, simulations using network predictions are compared against those using the original geometric signed distance evaluation. As shown in Fig.~\ref{fig:collapse_circular_NN}, the two approaches produce closely matching final deposit morphologies and AOR for circular grains, indicating that the FiLM-SDF surrogate preserves the essential grain-scale interactions and macroscopic evolution.

\begin{figure}[!htbp]
    \setlength{\abovecaptionskip}{0cm}
    \setlength{\belowcaptionskip}{0cm}
    \centering
    \includegraphics[width=0.5\linewidth]{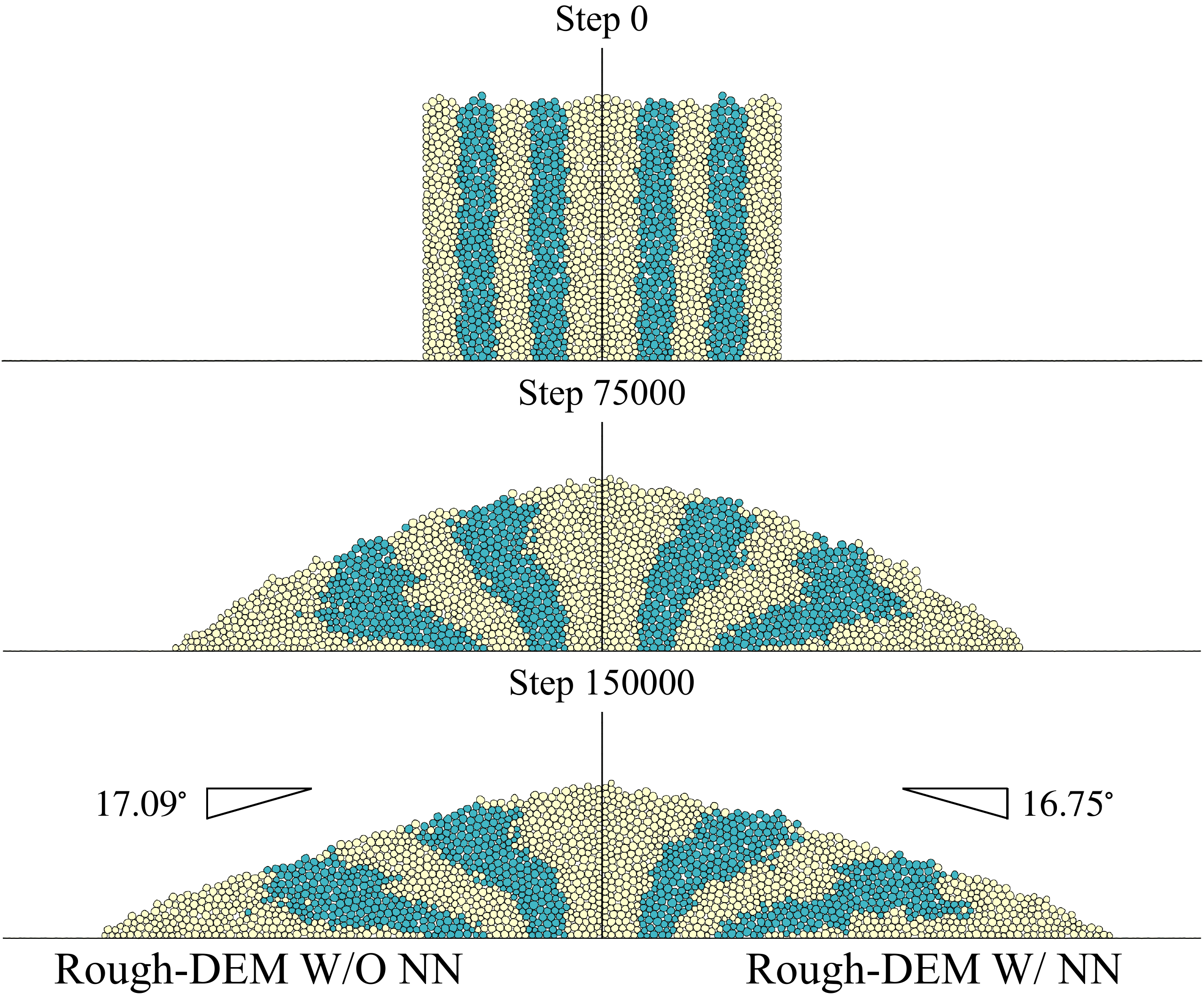}
    \caption{Comparison of column collapse simulations of irregularly shaped grains using the proposed Rough-DEM model without (left) and with (right) the NN surrogate. The final repose angles are indicated.}
    \label{fig:collapse_irregular_NN}
\end{figure}

Building on this, we further extend the comparison to irregularly shaped grains, generated by perturbing the control points $\vP_i$ of the circular base geometry with random offsets uniformly sampled from $[-0.12,0.12]\,\mathrm{mm}$. Owing to the geometry-conditioned architecture of the FiLM-SDF network, these irregular geometries can be accommodated without retraining, simply by supplying the corresponding control-point perturbation vectors as input. As shown in Fig.~\ref{fig:collapse_irregular_NN}, the NN-accelerated and original Rough-DEM simulations again produce closely matching final deposits and AORs, demonstrating the accuracy and generalizability of the surrogate across different grain geometries.

\subsection{Scaling Performance of Neural Network Acceleration}
\label{subsec:acceleration_scaling}
\begin{figure}[!htbp]
    \setlength{\abovecaptionskip}{0cm}
    \setlength{\belowcaptionskip}{0cm}
    \centering
    \includegraphics[width=0.85\linewidth]{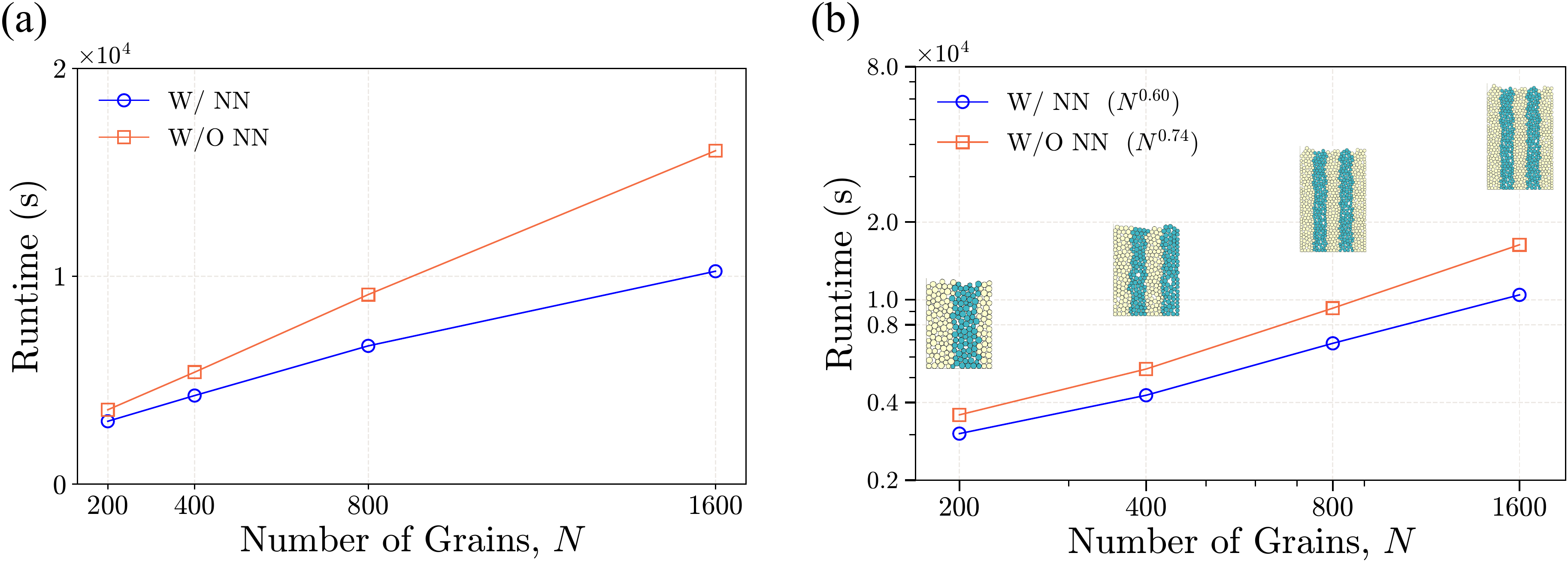}
    \caption{Total runtime of the Rough-DEM simulations, with and without the FiLM-SDF surrogate, for systems containing 200, 400, 800, and 1600 grains on an NVIDIA RTX 6000 GPU: (a) runtime as a function of the number of grains; (b) corresponding log-log representation with fitted scaling exponents.}
    \label{fig:scaling_performance}
\end{figure}

To quantify the computational benefits of introducing the NN and the associated lookup-table strategy, we compare the runtimes of the Rough-DEM simulations with and without NN acceleration over increasing system sizes. Specifically, each simulation is run for 100,000 frames on an NVIDIA RTX 6000 GPU, with systems comprising 200, 400, 800, and 1600 grains. As shown in Fig.~\ref{fig:scaling_performance}(a), the total runtime increases with the number of grains for both approaches, while the NN-accelerated implementation consistently achieves lower computational costs. The resulting speedup increases from approximately $1.2\times$ for 200 grains to $1.6\times$ for 1600 grains, indicating a growing performance advantage as the system size increases.

The scaling behavior is further examined on a log-log scale in Fig.~\ref{fig:scaling_performance}(b). Power-law fits give approximately $T\propto N^{0.60}$ for the NN-accelerated model and $T\propto N^{0.74}$ for the model without NN acceleration. Both implementations exhibit sublinear runtime growth over the investigated range of system sizes, with the lower exponent of the NN-accelerated model indicating a more favorable empirical scaling behavior. This indicates that introducing the surrogate does not impose an additional computational scaling penalty and instead provides a growing efficiency advantage within the tested system sizes.

Importantly, this acceleration is achieved by replacing only the computationally expensive roughness-scale signed distance evaluation, while retaining the original contact mechanics and time integration scheme. Unlike approaches that replace a larger portion of a computational granular dynamics pipeline with neural surrogates (see, e.g.,~\cite{lai2022machine,huang2024machine}), the proposed strategy targets a specific computational bottleneck without altering the underlying Rough-DEM formulation. The results therefore demonstrate that the FiLM-SDF surrogate and lookup-table strategy can improve computational efficiency while maintaining favorable scalability as the granular system size increases. Finally, we believe that Rough-DEM's scaling performance may be further improved by exploring alternative neural network structures and their utilization in representing the geometry of both the base shape and the surface roughness of a grain (see for example~\cite{park2019deepsdf}), both of which could be especially useful for three dimensions. We leave these investigations as part of our future work. 

\section{Conclusions and Discussions}
\label{sec:conclusion_outlook}
In this work, we propose an alternative framework, termed Rough-DEM, for modeling granular systems in which small-scale surface textures are explicitly resolved. Rather than prescribing an empirical friction law, we let macroscopic friction-like behavior emerge naturally from frictionless normal interactions between surface asperities. We begin with grains characterized by simple sinusoidal roughness, where the roughness amplitude $h$ and wavelength $\lambda$ serve as physically interpretable geometric parameters in place of tangential contact stiffness $k_t$ and the Coulomb friction coefficient $\mu_\text{p}$, without changing the overall model complexity. Through both two-body and many-body simulations, we demonstrate that Rough-DEM can reproduce typical predictions from classical DEM, suggesting that substantial tangential resistance can arise solely from normal contact interactions between (even simple, sinusoidal) rough surfaces. More importantly, even by tuning the simple sinusoidal surface roughness geometry, Rough-DEM can go beyond the confines of classical DEM by capturing strong geometric interlocking mechanisms within granular materials without pushing the friction coefficient to artificially large values, such as $\mu_\text{p}\geq 1$ (which may introduce additional empirical ambiguity into the model).

Rough-DEM also allows for modeling more complex surface roughness through a Fourier-series description, enabling the construction of geometries that more closely resemble ``arbitrary'' and self-similar rough surfaces encountered in real granular materials. We further validate this representation through column collapse simulations, demonstrating that an ``arbitrary'' roughness design can likewise be mapped onto an effective macroscopic friction coefficient in a collective sense, as quantified primarily by the resulting AOR.

The increased geometric complexity associated with explicitly resolving surface roughness, however, introduces a substantial computational burden in many-body contact evaluation. To alleviate this cost, we employ an acceleration strategy that combines a FE$^2$-like contact calculation algorithm and a FiLM-SDF surrogate that are parallelized on GPU, making roughness-resolved simulations computationally tractable at larger scales.

Taken together, Rough-DEM represents a paradigm shift for DEM-based modeling by incorporating physical insights from tribology and leveraging technical advances in AI and scientific computing. In particular, Rough-DEM establishes surface geometry as a physically interpretable mechanism for generating friction-like behavior across scales, from individual contacts to collective granular dynamics, without relying on an explicit tangential friction law. This perspective provides a direct link between microscale roughness and macroscale resistance and offers a route toward granular models parameterized by physically meaningful geometric descriptors rather than purely phenomenological friction parameters.

Looking ahead, we outline several promising directions for future research building on the present work. The first direction is the most immediate: extending Rough-DEM to 3D and optimizing its implementation, for which AI agents may become very helpful. By simulating interactions among 3D rough grains and comparing numerical predictions with experiments on 3D-printed, surface-tailored grains, we can further assess and validate the role of explicitly resolved surface geometry in reproducing effective frictional behavior in granular systems. In addition, and further down the road, Rough-DEM could in principle be readily integrated\footnote{There may be technical challenges associated with numerical resolution in this regard, but challenges are also opportunities.} with other (often continuum-level) numerical techniques towards high-fidelity multi-scale (e.g., with MPM~\cite{yue2018hybrid,liu2018coupling,chantharayukhonthorn2025hybrid} or FEM~\cite{andrade2009multiscale,andrade2011multiscale,guo2014coupled}) and multi-physics (e.g., with CFD~\cite{zhao2013coupled,wang2019semi}, LBM~\cite{cook2004direct,feng2007coupled}, or SPH~\cite{potapov2001liquid,liu2022sph}) simulators\footnote{The referenced publications here are by no means comprehensive and should only be treated solely as a high-level reflection of how DEM has been utilized within the computational mechanics community.}.

A second, more mechanism-based direction is to incorporate contact-induced elastic and possibly plastic deformation of roughness asperities, for example through finite element analysis, to investigate under what conditions these asperity-level mechanical deformations remain negligible (as is the case assumed in Rough-DEM and classical DEM) or instead become important. Such a framework could be used to establish quantitative relationships among roughness geometry, statistical roughness characteristics, material properties, and the resulting macroscopic effective friction response, and it may provide a bottom-up route toward a physics-based yet simple-enough friction model for practical applications. This is clearly a much broader problem. Friction does not arise solely from the mechanical deformation and interlocking of surface asperities, but may also involve adhesion, wear, rupture, interfacial chemistry or tribochemistry, and other microscale mechanisms \cite{vakis2018modeling}. Nevertheless, a central implication of the present work is that reproducing the dominant macroscopic response is possible by a ``tactically simplified'' and implicit representation of the underlying physical mechanisms, without explicitly incorporating---in a direct, additive (and possibly coupled and hence sophisticated) manner---every microscopic contribution. Taken together, progress in this direction would also make it possible to standardize the input roughness parametrization---just as classical DEM selects $k_t$ and $\mu_\text{p}$---for a given modeling problem at hand.

A third direction is more application-oriented. Because surface roughness is represented explicitly in Rough-DEM, it opens the possibility of constructing a differentiable inverse-design framework in which the roughness geometry of a contact interface is optimized to achieve a prescribed macroscopic response, with potential applications to designing on-demand the rheology of dense (non-Brownian) frictional suspensions ~\cite{hsiao2019experimental,singh2026jamming} and the mechanics of contact-driven architected metamaterials~\cite{wang2021structured,lu2024role,zhou20253d}. A central challenge lies in the nonsmooth nature of contact dynamics: the temporal formation and loss of contacts, changes in contact network, and stick-to-slide transitions can introduce discontinuities or poorly defined gradients. Nevertheless, differentiable contact formulations based on smooth regularization, soft-contact approximations, surrogate gradients, or related techniques may provide a practical route toward gradient-based optimization of surface geometry \cite{zhong2023improving, schwarke2024learning}.

\section*{CRediT authorship contribution statement}

\textbf{Xiwei Pan}: Software, Formal analysis, Conceptualization, Investigation, Methodology, Visualization, Validation, Writing - Original Draft, Writing - Review \& Editing. \textbf{Liuchi Li}: Conceptualization, Supervision, Investigation, Methodology, Funding acquisition, Project administration, Writing - Review \& Editing.

\section*{Declaration of competing interest}
The authors declare that they have no known competing financial interests or personal relationships that could have appeared to
influence the work reported in this paper.

\section*{Data availability}
Data will be made available upon reasonable request.

\section*{Code availability}
The source code of Rough-DEM will be made publicly available on GitHub upon the acceptance of publication of this work.

%%%%%%%%%%%%%%%%%%%%%%%%%%%%%%%%%%%%%%%%%%%%%%%%%%%
\section*{Acknowledgments}
The authors are pleased to acknowledge that the work reported on in this paper was substantially performed using the Princeton Research Computing resources at Princeton University. Princeton Research Computing is a consortium of groups including the Princeton Institute for Computational Science and Engineering (PICSciE) and Research Computing at Princeton University. L. L. gratefully acknowledges Princeton University and the Department of Civil and Environmental Engineering for financial support provided through the start-up funding.

\section*{Declaration on the Use of Generative AI}
The authors used ChatGPT and Claude, large language models developed by OpenAI and Anthropic, respectively, to assist with English language editing and grammar refinement. The scientific content, technical interpretations, and conclusions are solely the responsibility of the authors.




\bibliographystyle{elsarticle-num} 
\bibliography{references}

\end{document}